\documentstyle[epsfig,12pt]{article}
\begin{document}
\renewcommand{\thepage}{ \arabic{page}}
\def\ab#1{\renewcommand{\theequation}{#1}}
 \def\sous#1{\addtocounter{equation}{-#1} }
\def\normal{\renewcommand{\theequation}{\arabic{equation}} }
\renewcommand{\theequation}{\thesection.\arabic{equation}}
\begin{quote}
\raggedleft
PM/97-26 \\
GDR-S-002 \\
\end{quote}

\vspace*{3cm}
\begin{center}
{\bf Novel Electroweak Symmetry Breaking Conditions \\
\vspace{-0.3cm}
 From Quantum Effects In The MSSM}\footnote{supported
in part by EC contract CHRX-CT94-0579}
\end{center}
\setlength\baselineskip{12pt}
\begin{center}
Christophe LE MOU\"EL and Gilbert MOULTAKA\footnote{
Postal address as above. E-mail: moultaka@lpm.univ-montp2.fr,
phone: (33) 04.67.14.35.53,\\
fax: (33) 04.67.54.48.50}
\end{center}
\vspace{-0.5cm}
\begin{center}
{\it Physique Math\'ematique et Th\'eorique\\
UMR-CNRS,\\
Universit\'e Montpellier II, F34095 Montpellier Cedex 5, France}
\end{center}

\vspace{0.5cm}
\begin{abstract}

We present, in the context of the Minimal Supersymmetric Standard Model,
a detailed one-loop {\it analytic} study  of the 
minimization conditions of the effective potential in the Higgs sector.
 Special emphasis is put on the role played by $Str M^4$ in 
the determination of the electroweak symmetry breaking conditions, 
where first and second order derivatives of the effective potential 
are systematically taken into account. 
Novel, necessary (and sufficient in
the Higgs sector) {\sl model-independent} constraints,
are thus obtained analytically, 
leading to new theoretical lower and upper bounds on $\tan \beta$.
Although fully model-independent, these bounds are found to be
much more restrictive than the existing model-dependent ones! 
A first illustration is given in the context of a SUGRA-GUT motivated scenario.

\end{abstract}
\vfill
\newpage
\setlength\baselineskip{18pt}

\begin{center}
\section{Introduction}
\end{center}

Unraveling the origin of the electroweak symmetry breaking (EWSB) remains so far a piece of
theoretical challenge, and justifies a wealth of experimental endeavor. 
Even though 
(in a ``theory-of-everything'' state of mind) one assumes that
it is ultimately the dynamics at tremendously high
energy scales (GUT or Planck or
even higher) that should account for the fact that
the electroweak vacuum is not (gauge) symmetric, a central question remains:\\
how would this dynamics effectively translate, 
over tens of orders of magnitude, down to the electroweak scale?

Phenomenological models of supersymmetry offer plausible scenarios for
EWSB \cite{EWB} triggered by supersymmetry breaking, 
the latter being presumably connected
with dynamics at much higher scales \cite{hiddensector}, \cite{gaugemed}.
As is well-known, this {\it radiative} EWSB \cite{EWB} is made possible thanks 
to the 
heaviness of the top quark whose experimental discovery around $m_t\sim 180 GeV$
revived the interest in such models. 
Nonetheless, the breaking of the electroweak symmetry is not a matter of fact
in these scenarios,
in that one has still to enforce {\sl phenomenologically} a bounded 
from below potential and the absence of charge or color breaking minima 
\cite{casas}.
This is clear indication that the scheme is not yet dynamically complete, 
and one might ask whether a mechanism can be found which would naturally 
``lock'' the spontaneous symmetry breaking in the electroweak sector of the 
theory, at the relevant energy scale. 
The answer to this is obviously related to the general question asked above, and is 
out of the scope of the present paper. It will constitute, however, 
a general motivation for our study. Indeed, an important prerequisite to all 
this is to improve
our understanding of the structure of the effective potential of the 
Minimal Supersymmetric Standard Model (MSSM) and with it the sufficiency
and necessity of the phenomenological conditions for radiative EWSB.
One would like to assess as much as possible the implications of the
softly broken supersymmetry on the structure of the minima, before bringing
in more model-dependent information. For instance, as we will show later on,
some constraints involving $\tan \beta$ and usually quoted in a
model-dependent context, do have fully model-independent counterparts
in the MSSM, which are even more stringent!

Getting fairly reasonable approximations of the effective potential
is not always an easy task, for many reasons.
Many papers have been devoted to these aspects in the MSSM (see for instance
\cite{spectrum, gamb}), however often with
the main concern of reproducing a more or less quantitative pattern of the mass
spectrum, at the expense of making some rough assumptions about what the 
radiative EWSB conditions actually are, and about the scale at which they
are realized. This is of course fully justified, but only as a first 
approximation, as recognized by some authors \cite{hon}. The aim of the
present paper is {\sl not} to provide yet another tentative mass spectrum in the
context of some approximations, but rather to fit into a more theoretical study
of the validity of some of the underlying assumptions. Let us first list, for 
further reference, the features that should be kept in mind in this context:

{\sl i)} The true effective potential $V_{eff}$ should be renormalization scale invariant.
The renormalization group (RG) improved effective potential 
$\overline{V}_{eff}$
can miss large logarithms when there are more than one (mass) scale in the 
theory, as is the case in the MSSM. If not properly treated, the scale at which
leading logs are actually resummed becomes fuzzy and with it the approximate
scale invariance of $\overline{V}_{eff}$. {\sl ii)} The only gauge parameter independent 
and physically 
relevant values of $V_{eff}$ are those at stationary points. {\sl iii)}
The ``effective potential'' computed from the 1PI Green's functions at
vanishing external momenta, can suffer from unwanted concavities
( the true effective potential being always convex), precisely in the region
where spontaneous symmetry breaking occurs. {\sl iv)} A local minimum is 
characterized by both $1^{st}$ {\it and} $2^{nd}$ order derivatives 
with respect to {\it all} 
scalar fields in the theory. {\sl v)} Non logarithmic contributions to 
$V_{eff}$ at a given loop order can alter the qualitative picture
of the ``tree-level improved'' effective potential.\\
Another [perhaps less stressed] feature in connection with {\sl iv)},
is that what is usually called 
``the condition of electroweak symmetry breaking'' 
and used as such in the literature, has to do,
strictly speaking, only with the requirement of a stationary point 
of the effective potential in the Higgs directions. 
This obviously calls for an immediate explanation. In this paper
we will show, in connection with {\sl iv)} and {\sl v)}, 
 that while the stationarity condition is sufficient, as far
as the tree-level and ``tree-level improved'' effective potential are
concerned, due to a non-trivial connection between $1^{st}$ and $2^{nd}$
order derivatives, it is no more the case when finite (non-logarithmic)
one-loop effects are taken into account. 
Naively, one would argue that such effects
cannot change much of the qualitative pattern of the tree-level minima.
The fact is that, even if the one-loop effective potential
is expected to differ, point-to-point, only perturbatively from its tree-level
value, it is still possible that its shape be {\sl locally} modified in such 
a way that new local minima (or at least stationary points) appear. 
This fact will bring in new
supplementary conditions to the usual EWSB ones, determined analytically
in a general context. These conditions will lead in turn to model-independent
and calculable exclusion domains of $\tan \beta$, to compare with some existing
model-dependent ones. Furthermore, one will have to assess the necessity
and sufficiency for the new conditions. For this we will have to revisit 
briefly the question of charge breaking in the Higgs sector.\\

Another relevant aspect which we will consider, has to do with the gauge-fixing
dependence of the derived conditions. We will give a simple argument in favour 
of the gauge-fixing independence, in the sense that the corresponding equations
(and inequalities)
remain satisfied even if there is some gauge-fixing dependence on either side
of these equations, that is they are gauge-fixing ``covariant''.\\

The rest of the paper will be organized as follows. In section 2 we recall the 
main features of the effective potential in the MSSM and present some of the
assumptions we make throughout the paper and conventions we use. Section 3 
describes what we call the ``tree-level logic'' for the EWSB conditions. In
section 4 we give an expression for $Str M^4$ in the Higgs sector,
derive the new EWSB conditions at one-loop order in a fully analytic form,
and show why the ``tree-level logic'' does not generally hold. 
We then derive the model-independent theoretical bounds on $\tan \beta$.
The contents of
this section represent the main results of the paper.
In section 5 we discuss the meaning of our constraints and how they compare to 
the existing ones.
The question of necessity and sufficiency of our conditions
is addressed in section 6.   
Section 7 is devoted to the question of gauge-fixing dependence.
A first numerical illustration is given in section 8, in the context of 
SUGRA-GUT scenarios. The conclusions are given in section 9, 
and some technical material and proofs
relegated to the appendix. For a quick summary of the issues presented in
this paper the reader is referred to \cite{moriond}. This is part of a 
thorough analytic study of the
complete one-loop effective potential in the MSSM, where symbolic computation
is heavily used \cite{christophe}.

In this paper the word ``stationarity'' will refer to the vanishing of
the $1^{st}$ order derivatives of the effective potential, and the
word ``stability'' to the requirement that the eigenvalues of the $2^{nd}$ 
order derivatives should be positive.

\newpage
\begin{center}
\section{Approximations and notations}
\setcounter{equation}{0}
\end{center}
Our starting point will be the full fledged MSSM with {\sl a priori} no model
dependent assumptions whatsoever. The aim of the study is to assess the
power of the only assumption of minimal supersymmetry in constraining the
various free parameters, once electroweak symmetry breaking is required.
It is nonetheless unavoidable, for practical purposes, to take up a definite 
point of view,
as far as some approximations, usually made in a model-dependent
context, are concerned. For instance, regarding point {\sl i)} of the
previous section, it has been emphasized in the literature that the 
mere replacement of the free parameters by their runnings
in the tree-level effective potential, 
is actually a poor approximation \cite{gamb}. A prominent drawback
of this approximation is the rather high sensitivity of the ``improved''
effective potential to the renormalization scale. It was then observed that
the same procedure applied to the one-loop effective potential leads to
a better approximation, in the sense that the results are numerically much
more stable against the change of renormalization scale. \cite{gamb, decarlos}.

One should, nevertheless, keep in mind that these approximations cannot be
put in a neat context of leading- ( sub-leading-, etc...) log approximations
anyway,
as long as the case of many different mass and field scales is not properly
handled \cite{multiscale}. For instance,  there are still
regions where the vev's still suffer a pronounced scale dependence,  
even to one-loop order \cite{bando}. In a sense, this should not
come as a surprise since not all the logs can be properly resummed this way. 
For this reason, and since our aim is to study
the effects of the non-logarithmic one-loop contributions, we will be
contented throughout with the tree-level improved effective potential
approximation. 
 
The 1-loop effective potential has the well-known form \cite{CW}
\begin{equation}
V= V_{tree} + \frac{\hbar}{64 \pi^2} Str[ M^4 (Log \frac{M^2}{\mu_R^2} -
3/2) ] \label{EP}
\end{equation}

in the $\overline{MS}$ scheme. Here $\mu_R$ denotes the renormalization scale, 
$M^2$ the field dependent squared mass matrix of the scalar 
fields,
and $Str[...] \equiv \sum_{spin} (-1)^{2 s} (2 s + 1) (...)_s $, where the sum
runs  
over gauge boson, fermion and scalar contributions.
$V_{tree}$ is the tree-level MSSM potential \cite{susy}.
As stated above, we will assume that the logarithms in Eq.(\ref{EP})
are reabsorbed in the running of all the parameters
in $\overline{V}_{tree}(\mu_R^2$). This amounts to the rough
assumption of neglecting altogether all field and mass scale differences
in the logs. 
The effective potential takes then the form
\begin{equation}
 V= \overline{V}_{tree}(\mu_R^2) + \frac{\hbar}{64 \pi^2} (-3/2) Str M^4 
\label{Poteff}
\end{equation}

Here $\overline{V}_{tree}(\mu_R^2)$ is obtained from $V_{tree}$ by replacing
all the tree-level quantities by their running counterparts.

In our convention, the  superpotential reads
\begin{equation}
W={\sum_{i,j=gen} Y^u_{ij} {\tilde {u}}_{R}^{i} { H}_2.{\tilde{ Q}}^j+Y^d_{ij} 
{\tilde{ d}}_{R}^{i} {{ H}}_1.{\tilde{ Q}}^j+
       Y^l_{ij} {\tilde{ l}}_{R}^{i} {{ H}}_2.{\tilde{ L}}^j+\mu {{ H}}_2.{{ H}}_1}
\end{equation}
where ${\tilde{ u}}_{R}$, ${\tilde{ d}}_{R}$, ${\tilde{ Q}}$ (resp. ${\tilde{ l}}_{R}$, 
${\tilde{ L}}$) denote squark (resp. slepton) right and left-handed  scalar fields, and 
${{ H}}_1$, ${{ H}}_2$
the Higgs scalar fields. The product between $SU(2)_{L}$ doublets reads
$H_.Q \equiv \epsilon_{i j} H^i Q^j$ where $i, j$ are $SU(2)_L$ indices.\\
The tree level potential $V_{tree}$ is the sum of the so-called F- and D-terms, 
where F-terms come from the superpotential 
through derivatives with respect to all scalar fields $\phi_{a}$
\begin{equation}
V_{F}={\sum_{a} |W^{a}|^2} \ \ , \ \ \ W^{a} = {\frac{\partial{W}}{\partial{ \phi_a}}}
\end{equation}
and D-terms corresponding to respectively $U(1)_{Y}$, $SU(2)_L$, and $SU(3)_C$ 
gauge symmetries are given by
\begin{equation}
V_{D}= D_{1}D_{1}+ D_{2}D_{2}+ D_{3}D_{3}
\end{equation}
with
\begin{eqnarray}
D_1&=&g_1^2 [{\sum_{i=gen}{({\frac{1}{6}} {\tilde{Q}}_i^{\dagger}{\tilde{Q}}_i  
-{\frac{1}{2}} {\tilde{L}}_i^{\dagger} {\tilde{L}}_i -{\frac{2}{3}} 
{\tilde{u}}_{R_i}^{\dagger}{\tilde{u}}_{R_i} +
                      {\frac{1}{3}} {\tilde{d}}_{R_i}^{\dagger}{\tilde{d}}_{R_i}+  
{{\tilde{l}}_{R_i}^{\dagger}{{\tilde{l}}_{R_i} )}}}} \nonumber \\
&&      + {\frac{1}{2}}H_2^{\dagger} H_2-{\frac{1}{2}}H_1^{\dagger} H_1] \\
D_2&=&g_2^2 [{\sum_{i=gen}{ ({\tilde{Q}}_i^{\dagger} {\frac{\vec{\sigma}}{2}} 
{\tilde{Q}}_i + 
{\tilde{L}}_i^{\dagger} {\frac{\vec{\sigma}}{2}} {\tilde{L}}_i)+ 
H_2^{\dagger} {\frac{\vec{\sigma}}{2}}{H_2}
      +H_1^{\dagger} {\frac{\vec{\sigma}}{2}}{H_1}] }} \\
D_3&=&g_3^2 {\sum_{i=gen}{ [{\tilde{Q}}_i^{\dagger} {\frac{\vec{\lambda}}{2}} 
{\tilde{Q}}_i- 
{\tilde{u}}_{R_i}^{\dagger} {\frac{\vec{\lambda^{*}}}{2}} {\tilde{u}}_{R_i}-
         {\tilde{d}}_{R_i}^{\dagger}{\frac{\vec{\lambda^{*}}}{2}} {{\tilde{d}}_{R_i}]}}}
\end{eqnarray}

Here $g_1, g_2, g_3$ denote the three gauge couplings,  
 $({\sigma_{k}})_{k=1,..,3}$ and $( {\lambda}_{k})_{k=1,..,8}$ 
the  Pauli and Gell-Mann matrices.\\
To calculate $Str M^4$, we must also specify our conventions for the 
supersymmetric couplings of the scalar-fermion sector. 
Following the conventions of 
\cite{wessb} , with two-component spinors, this part of the Lagrangian 
contains on one hand a purely chiral sector
\begin{equation}
{\cal L}_{chir.}=-{\frac{1}{2}}{\sum_{a,b} { {W^{ab}} \psi_{a} \psi_{b}}}+h.c \ \
 , \ \  {W^{ab}}={\frac{\partial^2{W}}{\partial{ \phi_a \partial{ \phi_b}}}}
\end{equation}
where $\psi_{a}$ is the supersymmetric fermionic partner of $\phi_{a}$, and on 
the other hand a mixed chiral-vector sector, which yields in component form 
\begin{equation}
{\cal L}_{mix}=i{\sqrt{2}} {\sum_{A,a}(D_{A})^{a} \psi_{a} \lambda_{A}}+h.c \ \ , 
\ \, (D_{A})^{a} = {\frac{\partial{D_{A}}}{\partial{ \phi_a}}}
\end{equation}
The index A denotes the gauge group with which  the gaugino $\lambda_{A}$  is associated.\\
The soft scalar supersymmetry breaking terms take the form
\begin{eqnarray}
Vsoft&=& m^2_{H_2} H_2^{\dagger} H_2+m^2_{H_1}  H_1^{\dagger} H_1-B \mu (H_1.H_2+h.c) 
\nonumber \\
&& +{\sum_{i=gen} m^2_{{\tilde {Q}},i} {\tilde{Q}}_i^{\dagger}{\tilde{Q}}_i+
m^2_{{\tilde{ L}},i} {\tilde{L}}_i^{\dagger} {\tilde{L}}_i +
         m^2_{ {\tilde{u}},i} |{\tilde{u}}_{R_i}|^2+m^2_{ {\tilde{d}},i} |{\tilde{d}}_{R_i}|^2+  
      m^2_{{\tilde{l}},i} | {\tilde{l}}_{R_i}|^2}    \nonumber \\      
&& +{\sum_{i,j=gen} {(A^u_{ij} Y^u_{ij}  {\tilde{u}}_{R_i} H_2. {\tilde{Q}}_j+
A^d_{ij} Y^d_{ij}  {\tilde{d}}_{R_i} H_1.{\tilde{Q}}_j
+A^l_{ij} Y^l_{ij} {\tilde{l}}_{R_i} H_2.{\tilde{L}}_j+h.c)}}
\end{eqnarray}
Finally, the gaugino soft mass term reads
\begin{equation}
{\cal L}^{gaugino}_{mass}=-{\frac {1}{2}} {\sum_{i=1}^3 M_{i} \lambda_{i} \lambda_{i} + h.c}
\end{equation}
We are interested here in the Higgs directions 
in which case $V_{tree}$ takes the form,  
\begin{eqnarray} 
V_{tree}&=& m_1^2 H_1^{\dagger} H_1 + m_2^2 H_2^{\dagger} H_2 + m_3^2 (H_1.H_2 +
 h.c.) \nonumber \\
&&+ \frac{g^2}{8}  ( H_1^{\dagger} H_1 - H_2^{\dagger} H_2)^2 + \frac{g_2^2}{2} 
(H_1^{\dagger}  H_2) (H_2^{\dagger}  H_1) 
\label{vtree}
\end{eqnarray}

where $g^2 \equiv g_1^2 + g_2^2$, $m_1^2=m^2_{H_1}+\mu ^2$,  
$m_2^2=m^2_{H_2}+\mu ^2$, $m_3^2 =- B \mu $, and
\begin{eqnarray} 
H_1= \left( \begin{array}{c} 
               H_1^0 \\
               H_1^{-} \\
               \end{array} \right) \hspace{2cm}
H_2= \left( \begin{array}{c} 
               H_2^{+} \\
               H_2^0 \\
               \end{array} \right) 
\end{eqnarray}

\newpage
\begin{center}
\section{The ``tree-level logic''}
\setcounter{equation}{0}
\end{center}

In this section we will show, starting from the potential, Eq.(\ref{vtree}),
that 

the requirement of {\sl boundedness from below} of the potential,
implies that any 
gauge non-invariant {\sl stationary} point in the Higgs directions  
is {\sl necessarily a local minimum}, 
and that the existence of such a point guarantees the instability of the 
gauge symmetric ``vacuum''.

This interesting property is common knowledge, but, as far as we know, is 
seldom stated clearly, let alone shown explicitly. We do this here for several
reasons: First, to stress the fact that this property is not self-evident, but
rather a consequence of the
special form of Eq.(\ref{vtree}), {\sl i.e.} of (softly-broken) supersymmetry;
secondly, because it underlies the reason for which the usual EWSB conditions

\begin{eqnarray}
\frac{1}{2} M_Z^2 = \frac{\overline{m}_1^2 - \overline{m}_2^2 \tan^2 \beta}
{\tan^2 \beta -1 }  &,&
\sin 2\beta= \frac{-2 \overline{m}_3^2}{\overline{m}_1^2 + \overline{m}_2^2}
\label{EWSBcond} \\
\nonumber
\end{eqnarray}

used in the literature as {\sl minimization} conditions,
but which are strictly speaking only stationarity conditions, happen to be
safe, {\sl i.e.} necessary and sufficient,
as far as the effective potential has the form
of eq.(\ref{vtree}). We refer to this situation as the ``tree-level logic'',
even though it obviously holds as well in the tree-level-RG improved case; 
thirdly, to set the stage for the next section where
the one-loop case will be worked out in detail and shown to infringe the
tree-level logic.

 Since we are mainly interested
here in whether a stationary point in the electroweak direction is a local
minimum or not, one could argue that it is enough to choose from the start
the (electrically) neutral directions 

\begin{eqnarray} 
<H_1>= \frac{1}{\sqrt{2}}\left( \begin{array}{c} 
               v_1 \\
               0 \\
               \end{array} \right) \hspace{2cm}
<H_2>= \frac{1}{\sqrt{2}}\left( \begin{array}{c} 
               0 \\
               v_2 \\
               \end{array} \right)
\label{neutrdir} 
\end{eqnarray}

as the relevant dynamical variables.
However, this needs further justification. The doublets $H_1$, $H_2$ depend
each on four real-valued fields so that, as stated in the introduction
(point {\sl iv)}), one should determine $1^{st}$ and $2^{nd}$ order derivatives
of $V_{tree}$  with respect to the eight real fields, and then take
$H_1= <H_1>,  H_2= <H_2>$ \footnote{Note that one can use the $SU(2)$ gauge 
freedom to take away three of the eight fields, but the question remains 
essentially the same.}. Nonetheless, it turns out that all
$1^{st}$ order derivatives other than those with respect to 
the two neutral real components, vanish
trivially at $H_1= <H_1>,  H_2= <H_2>$. The two usual stationarity conditions
are then indeed the only ones to consider and can be readily obtained from
Eq.(\ref{vtree}) as

\begin{eqnarray}
m_1^2 v_1 + m_3^2 v_2 + \frac{g^2}{8} v_1 ( v_1^2 - v_2^2) &=& 0 \label{Dtree1}
\\
m_2^2 v_2 + m_3^2 v_1 + \frac{g^2}{8} v_2 ( v_2^2 - v_1^2) &=& 0 \label{Dtree2}
\\
\nonumber
\end{eqnarray}

On the contrary, the $2^{nd}$ order derivatives of the potential 
yield {\sl \`a priori} 
non trivial information in the eight field directions. The stability 
conditions are obtained using the invariants method (see appendix B).
Substituting for $m_1$ and $m_2$ using Eqs. (\ref{Dtree1}, \ref{Dtree2}) one finds
five stability conditions, 

\begin{eqnarray}
&& - (v_1^2 + v_2^2) \frac{m_3^2}{ v_1 v_2} \geq 0 \label{inv1} \\
&& - (v_1^2 - v_2^2)^2 \frac{g^2 m_3^2}{ 4 v_1 v_2} \geq 0  \label{inv2} \\
&&\frac{(v_1^2 + v_2^2) (- 4 m_3^2 + g^2 v_1 v_2)}{ 4 v_1 v_2} 
\geq 0 \label{inv3}\\
&& \frac{( - 4 m_3^2 + g_2^2 v_1 v_2)}{4 v_1 v_2} \geq 0 \;\;\;\;\;\;
\mbox{(twice)}
\label{inv4}\\
\nonumber
\end{eqnarray}
plus three zeros corresponding to the three goldstone degrees of freedom. 
Eqs.(\ref{Dtree1} - \ref{inv4}) are the necessary and sufficient conditions
for the existence of a {\sl local} minimum in the electroweak sector\footnote{
It should  be clear that they are not sufficient to ensure a {\sl global}
minimum, the latter necessitating the comparison of various local minima,
including those outside the Higgs directions.}.
We will show now that Eqs.(\ref{Dtree1}, \ref{Dtree2}) imply automatically 
Eqs.(\ref{inv1} - \ref{inv4}).

Let us first solve for $ \tan \beta\ (\equiv \frac{v_2}{v_1})$ in 
Eqs.(\ref{Dtree1}, \ref{Dtree2}).  One easily finds 
,

\begin{equation}
 \tan \beta = \frac{ - m_1^2 - m_2^2 \pm 
\sqrt{ (m_1^2 + m_2^2)^2 - 4 m_3^4}}{2 m_3^2}
\label{soltbeta} 
\end{equation}

It is worthwhile noting at this level that the existence of real solutions
in Eq.(\ref{soltbeta}) is guaranteed by the requirement of boundedness
from below of $V_{tree}$. The conditions of boundedness from below
involve necessarily the soft supersymmetry breaking parameters and
reside entirely in the would-be ``flat directions'' of the potential. 
The obvious reason for this is that in exact supersymmetry the potential is
positive semi-definite so that only explicit soft breaking terms can make
it unbounded from below. Furthermore, the directions along which the potential
can turn down negative first, when the soft terms are switched on, are those 
for which the supersymmetric potential is the closest to zero
at arbitrarily large fields. 
This happens generically\footnote{barring the occurrence of runaway 
vacua,}
for D-flat or F-flat (or both) directions. In the case of Eq.(\ref{vtree}),
the flat directions are D-flat and correspond to 
\begin{equation}
|H_1|=|H_2| \;\;\;\mbox{and}\;\;\; H_1^{\dagger} H_2 =0
\end{equation}
A representative direction is, (modulo a gauge transformation),
\begin{eqnarray} 
H_1= \left( \begin{array}{c} 
                H \\
               0 \\
               \end{array} \right) \hspace{2cm}
H_2= \left( \begin{array}{c} 
               0 \\
               \pm H \\
               \end{array} \right) 
\label{flatdir}
\end{eqnarray}
where H denotes any complex valued field.
$V_{tree}$ is quadratic in H along such directions, and the boundedness from
below conditions (BFB) read,

\begin{equation}
m_1^2 + m_2^2 \pm 2 m_3^2 \geq 0
\label{ufbtree}
\end{equation}

With the above conditions, $\tan \beta$ in Eq.(\ref{soltbeta}) is real valued.

More importantly, further inspection of Eq.(\ref{soltbeta}), 
in conjunction with Eq.(\ref{ufbtree}), tells us
that $\tan \beta$ and $m_3^2$ have opposite signs, that is
\begin{equation}
\frac{m_3^2}{v_1 v_2} \leq 0
\label{oldcond}
\end{equation}
A quick look at the stability conditions Eqs.(\ref{inv1} - \ref{inv4})
shows then that they are all {\sl automatically satisfied}.

On the other hand, substituting for
$m_1$ and $m_2$, one gets from Eqs(\ref{Dtree1},\ref{Dtree2}), 

\begin{equation}
m_1^2 m_2^2 - m_3^4=-\frac{1}{64} g^2 (v_1^2 - v_2^2)^2 (g^2 -8 m_3^2/(v_1 v_2))
\label{ssb}
\end{equation} 

so that Eq(\ref{oldcond}) implies readily 
\begin{equation}
m_1^2 m_2^2 - m_3^4 < 0
\label{ssb1}
\end{equation}
{\sl i.e., the potential is automatically unstable at 
zero vev's}.

Up to this level, we have used only partially the conditions assuring the
existence of a stationary point away from the origin. In particular,
 the existence of
$\tan \beta$ solutions, Eq.(\ref{soltbeta}), does not guarantee the reality
of the vev's, i.e. $v_1^2 , v_2^2 > 0$. To see this, let us extract from Eqs.(
\ref{Dtree1},\ref{Dtree2}) the equation complementary to Eq.(\ref{soltbeta}), namely,

\begin{equation}
u\equiv v_1 v_2= \frac{8 t}{g^2 ( 1 - t^4)} ( t^2 m_2^2 - m_1^2)
\label{solv1v2}
\end{equation}
where $t\equiv \tan \beta$. It is clear from the above equation that requiring
$t$ and $u$ to have the same sign, i.e. $v_1^2 , v_2^2 > 0$, will lead to
extra relations among the ${m_i}^2$'s. The discussion can be carried out in terms
of the sign of $m_2^2$ (assuming throughout $m_1^2>0$, as is generically the 
case in 
susy-GUT scenarios). If $m_2^2<0$, which is often the case at the electroweak
scale\footnote{ Remember that $m_2^2<0$ is only sufficient to ensure an 
unstable gauge-invariant vacuum, the necessary (and sufficient) condition
being Eq.(\ref{ssb1})}, then one has from Eq.(\ref{solv1v2}) and $ u t >0$
that $|\tan \beta| <1$ is  {\sl excluded}, and otherwise no extra conditions
on the ${m_i}^2$'s. If however $m_2^2>0$, then one can show,  given 
Eqs.(\ref{soltbeta}, \ref{solv1v2}) and the BFB Eq.(\ref{ufbtree}),
that the sufficient and necessary conditions for $u t >0$ are simply
\begin{equation}
m_1^2 m_2^2 - m_3^4 < 0 \;\;\mbox{and} \;\; m_1^2 > m_2^2 \;\;(\mbox{resp.}\;\;
m_1^2 < m_2^2) \;\; \mbox{for}\;\; |t|>1  \;\;(\mbox{resp.} \;\;|t|<1)
\label{treebound}
\end{equation}

The first of these two conditions is, however, already a consequence of
Eq.(\ref{soltbeta}) as we have shown above.

To summarize,
we have shown in this section that the requirement of boundedness from below
Eq.(\ref{ufbtree}) and the conditions of existence of a stationary point
at non-vanishing field values Eqs(\ref{Dtree1}, \ref{Dtree2}), 
imply by themselves that this point is a local minimum
Eqs.(\ref{inv1} - \ref{inv4}), and that the gauge symmetric point, $v_1=v_2=0$
is energetically unstable Eq.(\ref{ssb}). On the other hand, 
the discussion following Eq.(\ref{solv1v2}) is to remind us that even at 
tree-level, the model-independent EWSB conditions imply some constraints
on the allowed values of $\tan \beta$, depending on the relative magnitude
and/or signs of $m_1^2, m_2^2$. Of course, if one implements the information
coming from say a SUGRA-GUT scenario together with universality assumptions
then the above constraints become stricter as  $m_1^2, m_2^2$ run down from
their initial values in a well-defined way. In particular, theoretical
upper bounds are known to appear (\cite{giudice}). The main point of the next section
is to show that similar constraints obtain when one considers the one-loop
EWSB constraints, {\sl irrespective} of model-dependent assumptions. 

Finally a couple of remarks are in order.
 
\begin{itemize}
\item First, the fact that the second order
derivatives are automatically positive when the first order ones are vanishing
can of course be also seen from the tree-level MSSM relations
among the Higgs masses. Indeed the well-known relations

\begin{equation}
m^2_{h^0, H^0}=\frac{1}{2}(m_Z^2 + m_{A^0}^2 \mp \sqrt{(m_Z^2 + m_{A^0}^2)^2 
- 4 m_Z^2 m_{A^0}^2 \cos^2 2 \beta}\;\; ) 
\end{equation}
\begin{equation}
m^2_{H^\pm} = m_{A^0}^2 + m_W^2
\end{equation}

are such that all squared Higgs masses are automatically positive once
$m_A^2 \geq 0$. Furthermore, since 

$m_A^2 = - (v_1^2 +  v_2^2) \frac{m_3^2}{v_1 v_2}$,
the positivity of $m_A^2$ is itself a consequence of Eq.(\ref{oldcond}),
which in turn is implied by the vanishing of the first order derivatives.
It thus follows that the positivity of all the squared Higgs masses is
at tree-level a consequence of just the stationarity conditions.

\item {\sl sign of $\tan \beta$}: For later discussion, let us also recall
the reason why $\tan \beta$ can always be chosen positive at tree-level.
As already stated,
Eq.(\ref{soltbeta}) implies by itself that the two $\tan \beta$ solutions
have the same sign, opposite to that of $m_3^2$. The latter can always be
chosen negative in Eq.(\ref{vtree}) by properly choosing the relative
phase of $H_1$ and $H_2$. One can thus replace from the start $m_3^2$ by
$-|m_3^2|$ in Eq.(\ref{vtree}) and choose all the way long 
$\tan \beta >0$. [The choice of positive $\tan \beta$ 
will still be possible beyond tree-level, however in a more
tricky way. Indeed, as we will see in the next section, the signs of 
$\tan \beta$ and the one-loop corrected $m_3^2$ are no more directly connected
through the stationarity conditions].

\item {\sl The case $\tan \beta=1$}: Whatever the theoretical $\tan \beta$
exclusion domains turn out to be, we would like to stress that the value
$\tan \beta=1$ should be excluded. 
As can be easily seen from Eqs.(\ref{Dtree1},
\ref{Dtree2}), when $v_1=v_2$ (and $\neq 0$) one has necessarily 
$m_1^2 + m_2^2 + 2 m_3^2=0$. This implies that the tree-level potential
Eq.(\ref{vtree}) is vanishing everywhere in the neutral direction
Eq.(\ref{neutrdir}) so that there is no dynamically
preferred electroweak minimum. 
As we will see in the next section, although the degeneracy is lifted by the 
one-loop corrections, flatness remains in the $v_1=v_2$ direction in such
a way that the stationary point at $\tan \beta=1$ is not a minimum.    
 
\end{itemize}

\begin{center}
\section{$Str M^4$ and the 1-loop EWSB conditions}
\end{center}
\setcounter{equation}{0}
The natural question to ask, given the results of the previous section,
is whether the non-trivial connection between stationarity and local stability
just demonstrated, can hold beyond the tree-level. For one thing, it is obvious
that it does in the tree-level-RGE improved approximation, where the potential
retains its tree-level field dependence, provided that
multi-scale subtleties be momentarily evaded by assuming roughly 
one energy scale, and making the substitutions 
$m_i^2 \rightarrow \bar{m}_i^2(\mu_R^2)$, 
$g_i \rightarrow \bar{g}_i(\mu_R^2)$. 
[The running of the Higgs fields being quite small is usually disregarded. 
In any case the implicit $ \mu_R^2$ dependence on the fields
is ignored within the approximation, 
so that one can keep theses fields fixed for the purpose of minimization.]

Potentially new information would come from the study of the
non-logarithmic one-loop correction to the effective potential, which we turn
to now. Here we stick as before to the $H_1, H_2$ directions, and keep as much
as possible the discussion at a model-independent level. 

Putting together the contributions of the full-fledged MSSM, we find
the following form for the one-loop effective potential in the Landau gauge,
in the Higgs directions 
\footnote{the full analytic form of $Str M^4$ in all scalar field directions 
will be given elsewhere \cite{christophe}},
\begin{eqnarray}
V_{tree} +\kappa Str M^4&=& X_{m_1}^2 H_1^{\dagger} H_1 + X_{m_2}^2 
    H_2^{\dagger} H_2 + 
          X_{m_3}^2 (H_1.H_2 + h.c.)  \nonumber\\
         &&+ X ( H_1^{\dagger} H_1 - H_2^{\dagger} H_2)^2 +
        \tilde{\beta} |H_1^{\dagger}  H_2|^2 
+ \tilde{\alpha} ((H_1^{\dagger} H_1)^2 - (H_2^{\dagger} H_2)^2) + \Omega_0  
\nonumber\\
\label{vloop}
\end{eqnarray}
where

\begin{eqnarray}
\tilde{\alpha} &=& \frac{3}{2} \kappa g^2({Y_t}^2 - {Y_b}^2) \label{alpha}\\
\tilde{\beta}&=& \frac{g_2^2}{2} + \kappa g_2^2 ( g_1^2 + 5 g_2^2 - 6(Y_t^2 +
Y_b^2) ) \\
X &=& \frac{g^2}{8} + \kappa (g_1^2 g_2^2  +
   \frac{23 g_1^4 + 5 g_2^4}{4} - \frac{3}{2} g^2({Y_t}^2+ {Y_b}^2 ))  \\
\nonumber
\end{eqnarray}
\begin{eqnarray}
X_{m_1}^2 &=&  m_{H_d}^2 + \mu^2 + \kappa[ -4 g_1^2 M_1^2  +
3 g_2^2 (  m_{H_d}^2 - 3 \mu^2 -4 M_2^2)\nonumber \\
&& + 12 ( Y_b^2 ( A_b^2 + m_{\tilde{b_R}}^2 +
 m_{\tilde{T}}^2 ) + \mu^2 Y_t^2)  \nonumber \\
&&+ g_1^2 ( 3 m_{H_d}^2 - 2 m_{H_u}^2 - 3 \mu^2 \nonumber \\
&&- \sum_{i=generation} 2 (
m_{\tilde{d}_R, i}^2 +  m_{\tilde{l}_R, i}^2 + m_{\tilde{Q}, i}^2 -
m_{\tilde{L}, i}^2 - 2 m_{\tilde{u}_R, i}^2) ) ] \nonumber \\
\end{eqnarray}
\begin{eqnarray}
X_{m_2}^2 &=&  m_{H_u}^2 + \mu^2 + \kappa[ -4 g_1^2 M_1^2  +
3 g_2^2 (  m_{H_u}^2 - 3 \mu^2 - 4 M_2^2) \nonumber \\
&&+ 12 ( Y_t^2 ( A_t^2 + m_{\tilde{t_R}}^2 +
 m_{\tilde{T}}^2 ) + \mu^2 Y_b^2)  \nonumber \\
&& + g_1^2 ( 3 m_{H_u}^2 - 2 m_{H_d}^2 - 3 \mu^2 \nonumber \\
&&+ \sum_{i=generation} 2 (
m_{\tilde{d}_R, i}^2 +  m_{\tilde{l}_R, i}^2 + m_{\tilde{Q}, i}^2 -
m_{\tilde{L}, i}^2 - 2 m_{\tilde{u}_R, i}^2) ) ] \nonumber\\
\end{eqnarray}
\begin{eqnarray}
X_{m_3}^2 &=& -B \mu + \kappa \mu [ g_1^2 ( B + 4 M_1 ) + 
3 g_2^2 (  B + 4 M_2) \nonumber \\
&& ~~~~~~~~~~~~~~~~~~~~~~~~~ - 12 ( A_t Y_t^2 + A_b Y_b^2)]  \\
\kappa&=&(-\frac{3}{2})\frac{\hbar}{64\pi^2} \label{kappa}\\
g^2 &\equiv& g_1^2 + g_2^2 \\
\nonumber
\end{eqnarray}

The above expressions are exact, apart from the fact that we kept only the
top/bottom Yukawa couplings $Y_t$ and $Y_b$ for simplicity, the generalization
to the other Yukawa couplings being easy to implement,
and assumed diagonal and real Yukawa matrices.
The $X_{m_i}^2$'s are functions
of the various soft susy breaking masses and couplings, associated with
all squark-doublet-masses, $m_{\tilde{Q}, i}, m_{\tilde{T}}$, and singlet-
masses,
$m_{\tilde{u}_R, i}, m_{\tilde{d}_R, i}, m_{\tilde{b}_R}, m_{\tilde{t}_R}$,
all slepton-doublet-masses
, $m_{\tilde{L}, i}$ and singlets masses $m_{\tilde{l}_R, i}$
(no right-handed $\tilde{\nu}$), gaugino soft masses $M_1, M_2$ (no
gluino contributions at this level in the Higgs directions), Higgs soft masses 
$m_{H_1}^2, m_{H_2}^2$
and the supersymmetric $\mu$-term related to $m_1^2, m_2^2, m_2^3$ as defined
in section 2,      
and the soft trilinear couplings $A_t, A_b$.
Note also that $\Omega_0$ in Eq.(\ref{vloop}) is a field independent additive 
constant
depending exclusively on soft susy breaking terms.
It contributes to the cosmological constant and will be discarded throughout
\footnote{see however \cite{multiscale} for relevant issues}.
We stress that no model-dependent assumptions were used to establish
the above expressions. The results which will follow from them will thus be 
applicable either in the context of SUGRA-GUT scenarios, where eventually 
universality may be assumed \cite{hiddensector}, 
or in a gauge-mediated supersymmetry breaking context \cite{gaugemed}, 
or for that matter in any fully model-independent analysis.\\

{\it Boundedness from below}. The would-be flat directions remain 
(as expected)
the same as at tree-level, i.e.
\begin{equation}
|H_1|=|H_2| \;\;\;\mbox{and}\;\;\; H_1^{\dagger} H_2 =0
\end{equation}
and give the following one-loop condition for a bounded-from-below
potential, 
\begin{equation}
X_{m_1}^2 + X_{m_2}^2 \pm 2 X_{m_3}^2 \ge 0
\label{ufbloop}
\end{equation}
{\sl Stationarity conditions}. From Eq.(\ref{vloop}) we find the 
stationarity conditions with respect
to the 8 real components in $H_1$, $H_2$ at $H_1= <H_1>$, $H_2= <H_2>$.
As in the previous section, they boil down to two conditions,

\begin{eqnarray}
X_{m_1}^2 v_1 + X_{m_3}^2 v_2 + X v_1 ( v_1^2 - v_2^2) + \tilde{\alpha} v_1^3 = 0  &,&
X_{m_2}^2 v_2 + X_{m_3}^2 v_1 + X v_2 ( v_2^2 - v_1^2) - \tilde{\alpha} v_2^3 = 0  \nonumber \\
\label{EWSBnew}
\end{eqnarray}

It is suitable for the subsequent discussion to express these equations in
terms of $t\equiv \tan \beta$ and $u \equiv v_1 v_2$ as follows,

\begin{eqnarray}
X_{m_3}^2( \tilde{\alpha} - X) t^4  + ( \tilde{\alpha} X_{m_1}^2 -
X (X_{m_1}^2 + X_{m_2}^2 ) ) t^3  && \nonumber\\ 
+ ( \tilde{\alpha} X_{m_2}^2 
+ X (X_{m_1}^2 + X_{m_2}^2 ) ) t +  X_{m_3}^2 ( \tilde{\alpha} + X)
=0&& \label{eqtbeta}\\
\nonumber
\end{eqnarray}
\begin{eqnarray}
u = \frac{1}{\tilde{\alpha} (t^2 - 1)}
( X_{m_3}^2(t^2  + 1) + 
(X_{m_1}^2 + X_{m_2}^2 )t ) && \label{eqv1v2} \\
\nonumber
\end{eqnarray}

Let us pause here and make some important remarks for what will follow.
The variable $\tilde{\alpha}$ defined in Eq.(\ref{alpha}), which is thoroughly 
a one-loop contribution, will play a central role in the analysis leading to 
the new model-independent constraints on $\tan \beta$. The fact that it 
modifies, even though numerically slightly, the structure of the tree-level 
quartic terms in the effective potential  will generically 
invalidate the tree-level logic of the connection between $1^{st}$ and $2^{nd}$
order derivatives. Furthermore, $\tilde{\alpha}$ is proportional
to the {\sl difference} between up and down Yukawa couplings. This
will be at the origin of the $\tan \beta$ bounds involving 
$\frac{m_t}{m_b}$\footnote{
recall that we consider here just the top/bottom doublet}.  
Finally, $\tilde{\alpha}$  modifies drastically the $\tan \beta$ 
dependence of the stationarity conditions. Indeed, 
Eqs.(\ref{Dtree1}, \ref{Dtree2}) lead to an equation
quadratic in $\tan \beta$, while Eq.(\ref{eqtbeta}) is quartic, so that one 
should expect generically up to four possible real values for $\tan \beta$. 
As we will see, and already remarked in section 3, it would be erroneous then 
to discard from the start the negative valued ones.

{\sl Stability conditions}. We now go a little further and determine from 
Eq.(\ref{vloop}) the $2^{nd}$ order derivative matrix with respect to the 
eight real valued components of $H_1, H_2$, and extract from it 
(using the invariants method described appendix B), five stability
conditions, plus, as expected, three zeros corresponding to the goldstone
modes. We find, (after having eliminated $X_{m_1}^2, X_{m_2}^2$ through
Eqs.(\ref{EWSBnew})),

\begin{eqnarray}
&&-(v_1^2 + v_2^2) \frac{X_{m_3}^2}{v_1 v_2} \geq 0 \label{newinv1}
\end{eqnarray}
\begin{eqnarray}
&&2\tilde{\alpha} ( v_1^2 - v_2^2) + (v_1^2 + v_2^2)(2 X - \frac{X_{m_3}^2}{v_1 v_2} 
)\geq 0  \label{newinv2}
\end{eqnarray}
\begin{eqnarray}
&&-4 \tilde{\alpha}^2 v_1^2 v_2^2 + 2 (v_2^2 - v_1^2)[
(v_1^2 + v_2^2 ) \tilde{\alpha} - (v_2^2 - v_1^2) X] \frac{X_{m_3}^2}{v_1 v_2}
\geq 0 \label{newinv3}
\end{eqnarray}
\begin{eqnarray}
( -\frac{X_{m_3}^2}{v_1 v_2} + \tilde{\beta}) ( v_1^2 + v_2^2) \geq 0 
\;\; \mbox{(twice)}
\label{newinv4}
\end{eqnarray}

Eq.(\ref{newinv1}) is clearly equivalent to

\begin{equation}
\frac{X_{m_3}^2}{v_1 v_2} \leq 0
\label{newcond1}
\end{equation}
which is the straightforward one-loop generalization of Eq.(\ref{oldcond}).
Relying on the fact that the inequalities $ X \pm \tilde{\alpha} \ge 0$ and
$\tilde{\beta} \ge 0$ are always perturbatively satisfied, one shows
easily that Eqs.(\ref{newinv2}, \ref{newinv4}) hold as a consequence
of Eq.(\ref{newcond1}).

We are then left with Eq.(\ref{newinv3}). It can be easily shown that this 
equation is equivalent to 
\

\begin{equation}
 \tan^2\beta \leq t_{-} \mbox{\,or\,} \tan^2\beta \geq t_{+}
\label{newcond2} 
\end{equation}

where 
\begin{equation}
t_{\pm}= \frac{ \tilde{\alpha}^2 \frac{v_1 v_2}{X_{m_3}^2} - X  \mp
\sqrt{( X - \tilde{\alpha}^2 \frac{v_1 v_2}{X_{m_3}^2})^2 + \tilde{\alpha}^2-
X^2 }}{ \tilde{\alpha} - X }
\label{newcond2p}
\end{equation}
Note that in writing the above equations we assumed that Eq.(\ref{newcond1})
is verified, so that $t_{\pm}$ are both positive and satisfy
$t_{+} \ge t_{-} $, since $X$ and $X - \tilde{\alpha}$ are both perturbatively
positive.\\

To summarize, Eqs(\ref{newcond1}, \ref{newcond2}) constitute new stability
constraints, which are necessary
for EWSB, and should {\sl \`a priori} be superimposed on 
Eqs.(\ref{eqtbeta},\ref{eqv1v2}).
 
The question we address now is whether these stability conditions
are implicitly encoded in  
 the stationarity equations (\ref{EWSBnew}), as was the case
at the tree-level, or whether they are genuinely extra constraints.
The answer here is technically much less trivial, 
for two reasons:\\
-the rather involved form of the bounds $t_{\pm}$\\
-starting from the analytic solutions for Eq.(\ref{eqtbeta}),one can
exhibit cases, compatible with boundedness from below Eq.(\ref{ufbloop}), 
where  
$\tan \beta$ and $X_{m_3}^2$ {\sl do} have the same sign, contrary to what
was going on at tree-level.\\

{\bf $\frac{X_{m_3}^2}{v_1 v_2} \leq 0$ and $\tan \beta $ bounds}. 
As far as Eq.(\ref{newcond1}) is concerned,
the way out resides in Eq.(\ref{eqv1v2}) together with the requirement that
$u$ and $t$ should have the same sign; we will show that Eq.(\ref{eqv1v2})
encodes enough information to imply new lower and upper 
{\sl model-independent} bounds on $\tan \beta$. [The automatic validity of
Eq.(\ref{newcond1}) will persist only when 
$ 1 <|\tan \beta| < \frac{m_t}{m_b} $]. 

Let us give first a heuristic argument to illustrate how the interplay
between Eqs.(\ref{eqv1v2}), (\ref{newcond1}) and (\ref{alpha})
forbids $\tan\beta$
to be arbitrarily large or small.  
If the region $\tan \beta >>1$ were allowed, then Eq.(\ref{eqv1v2}) would 
behave like

\begin{equation}
 v_1 v_2 \sim \frac{X_{m_3}^2}{\tilde{\alpha}} 
\label{heur1}
\end{equation}
in that region. Taking into account Eq.(\ref{newcond1}) one then must have
$$ \tilde{\alpha} \leq 0$$
However, due to the form of $\tilde{\alpha}$, Eqs.(\ref{alpha}, \ref{kappa}), 
and to the fact that 
$ \tan \beta = \frac{Y_b}{Y_t} \frac{m_t}{m_b}$ one sees easily
that $ \tilde{\alpha} \leq 0$ and $\tan \beta$ arbitrarily large 
(i.e. $\frac{Y_b}{Y_t} >>1$) cannot
be simultaneously satisfied\footnote{
One should note that we rely here on the tree-level relations
$ m_t= \frac{Y_t v_2}{\sqrt{2}}, m_b = \frac{ Y_b v_1}{\sqrt{2}} $, which,
apart from leading log corrections in the running of $Y_t, Y_b, v_i$ ,
might suffer from some small corrections. Nonetheless, the qualitative form
of the constraint will not be altered} 
. We are thus lead to the conclusion that an
arbitrarily large $\tan \beta$ would lead to a contradiction, 
{\sl i.e. there should be a theoretical upper bound on $\tan \beta$}.\\
Similarly, in the region $|tan \beta| <<1 $ one has

\begin{equation} 
 v_1 v_2 \sim -\frac{X_{m_3}^2}{\tilde{\alpha}}
\label{heur2} 
\end{equation}
which, together with Eq.(\ref{newcond1}) implies that $\tilde{\alpha} \geq 0$,
the latter inequality being in contradiction with $|tan \beta| <<1 $.
Thus $\tan \beta$ cannot be arbitrarily small and {\sl a 
theoretical lower bound should exist}. To determine the actual
upper and lower bounds on $\tan \beta$ requires much more work. Here we only
state the results, assuming without loss of generality, that 
$\tan \beta$ is positive 
(and $X_{m_3}^2<0$). The complete proof and further comments, including the
correlation with the sign of $\tilde{\alpha}$, are deferred to
appendix A. We find the following bounds:

\begin{eqnarray}
&\mbox{\, if \,} \tan \beta > 1& \;\;\; \mbox{:} \;\;\;\tan \beta_{-} \leq 
\tan \beta \leq
\tan \beta_{+} \\
&& \nonumber \\
&\mbox{where}& \;\;\; \tan \beta_{-}=Min( T_{+}, \frac{m_t}{m_b})  \nonumber\\
&\mbox{and}& \;\;\;\tan \beta_{+}=Max( T_{+}, \frac{m_t}{m_b}) \nonumber\\
\nonumber
\end{eqnarray}
\begin{eqnarray}
&\!\!\!\!\!\!\!\!\!\!\!\!\!\!\!\!\!\!\!\!\mbox{\, if \,} \tan \beta < 1& \;\;\; 
\mbox{:} \;\;\; T_{-} \leq \tan\beta <1 \nonumber \\
\end{eqnarray}
 
\begin{equation}
\mbox{\,where\,\,}
T_{\pm} = \frac{ -X_{m_1}^2 - X_{m_2}^2 \mp \sqrt{(X_{m_1}^2 + X_{m_2}^2)^2 -
4  X_{m_3}^4 }}{2 X_{m_3}^2} \nonumber
\end{equation}

the $X_{m_i}^2$'s being given by Eqs.(4.6 - 4.8).
It is interesting to note at this level the possibility of
predicting the size (or even closing at will) the allowed window in the case
$\tan \beta > 1$.
This depends on whether $\frac{m_t}{m_b}$ is greater or smaller than $T_{+}$, 
(the latter being controlled by the values of the free parameters of the MSSM). 
For instance in the parameter space region
where 
\begin{equation}
X_{m_1}^2 + X_{m_2}^2 \simeq \frac{m_t}{m_b} |X_{m_3}^2|
\end{equation}

$\tan \beta$ is bound to lie very close to $\frac{m_t}{m_b}$ !

Finally, as we will also explicitly show in appendix A, the stability
condition Eq.(\ref{newcond1}) is {\sl not} a consequence of the
stationarity conditions, when 
$\tan \beta <1 \;\;\; \mbox{or} \;\;\;> \frac{m_t}{m_b} $
while it becomes so if $1 < \tan \beta < \frac{m_t}{m_b}$.
We find it rather noteworthy that the ``tree-level logic'' of section
2, holds only in the region of $\tan \beta$ which is favoured by
SUGRA-GUT scenarios, \cite{giudice} !\\

{\bf the $t_{\pm} $ bounds}. The $T_{\pm}$ are only part of the story.
It is of course also necessary to take into account the stability conditions
Eq.(\ref{newcond2}). When combining the bounds from
Eqs.(4.25,4.26) with those from Eq.(\ref{newcond2}) it will be
important to understand the relative
magnitudes of $\sqrt{t_{\pm}}$ as compared to $T_{\pm}$ and $\frac{m_t}{m_b}$,
in order to tell when the new $t_{\pm}$ bounds become relevant.
A complete answer to this would get too technical. Actually there is no 
generic trend for the relative magnitudes,
so that firm conclusions can be drawn only numerically and will depend on the
model assumptions. One can, however, show analytically that there is
no obstruction for $\sqrt{t_{+}}$ to lie within the interval
$[\tan \beta_{-}, \tan \beta_{+}]$. This means that the tree-level connection
between stationarity and stability is generally lost, even though it is
partially preserved for Eq.(\ref{newcond1}), as was stated at the end of the
previous paragraph. Hereafter we give, without proof, some of the salient 
properties 
of $t_{\pm}$.
 
Let us note first that the behaviour of $t_{\pm}$
in terms of $\zeta= \frac{\tilde{\alpha}^2 v_1 v_2}{X_{m_3}^2} $ is such
that $0\leq t_{-} \leq 1$ and $ t_{+} \geq 1$, whatever the actual magnitude of
$\zeta$ and the sign and magnitude of $\tilde{\alpha}$, provided that
Eq.(\ref{newcond1}) is verified, {\sl i.e.} $ \zeta \leq 0$. 
The proof of this is straightforward, and we will simply exemplify it
in Fig. 1; thus a calculable region around $\tan \beta =1$ is always excluded
by the stability constraint Eq.(\ref{newcond2}), the limiting value 
$\tan \beta =1$ being reached only when $\zeta = 0$.   
On the other hand,  $t_{\pm}$ depend on $t (\equiv \tan \beta)$
through $u (\equiv v_1 v_2)$, while the bounds $T_{\pm}$ do not. 
In view of Eq.(\ref{eqv1v2}), one should treat with some care the two
regimes $\tilde{\alpha} << 1$ and $t \sim 1$. In the regime 
$\tilde{\alpha} << 1$ and $t \neq 1$, $u$ is not singular,
due to the fact that the numerator of Eq.(\ref{eqv1v2}) is  $O(\tilde{\alpha})$
when Eq.(\ref{eqtbeta}) is taken into account. One finds, in the limit 
$\tilde{\alpha} \to 0$ four solutions, $-1, 1, T_{+}, T_{-}$. 
Excluding $|t| = 1$, the behaviour  of $u$ is then given by 
\begin{equation}
u \sim  \frac{ t}{ X ( 1 - t^4)} ( t^2 X_{m_2}^2 - X_{m_1}^2) + 
O(\tilde{\alpha})
\nonumber 
\label{limitu1}
\end{equation}
$$\mbox{where} \;\;\;\;\;\;\;\; t = T_{\pm}.$$ 
Note that this regime is relevant either when the one-loop correction is 
assumed to be small, or when $Y_t \sim Y_b$, {\sl i.e.} 
$\tan \beta \sim \frac{m_t}{m_b}$. In the first case, one should also make
the substitutions $ X_{m_i}^2 \to m_i^2, X \to \frac{g^2}{8}$, thus recovering
Eq.(\ref{solv1v2}).\\

The regime $t \simeq 1$ is somewhat more tricky. 
When $t \to 1$ one has simultaneously 

\begin{equation}
X_{m_1}^2 + X_{m_2}^2 + 2 X_{m_3}^2 \to 0 
\end{equation}
as can be easily seen from Eqs.(\ref{EWSBnew}), and the one-loop 
effective potential comes very close to flat in the $v_1 \sim v_2$ region.  
We thus parameterize this regime
by the small number $\epsilon= X_{m_1}^2 + X_{m_2}^2 + 2 X_{m_3}^2 $. 
To understand the effect of non-vanishing $\tilde{\alpha}$ we consider
the asymptotic region where $\epsilon << \tilde{\alpha}$. When $\epsilon$
is vanishing, Eq.(\ref{eqtbeta}) has only one solution $ t=1$. 
Barring the solution $ t= -1$ without loss of generality, we find three
positive solutions which are
away from 1 by at least $O(\sqrt{\tilde{\alpha}})$: 

\begin{eqnarray}
t_1 &=&  1 + \frac{ \epsilon}{ X_{m_2}^2 - X_{m_1}^2} + 
O(\epsilon^2,\tilde{\alpha})\\
t_2 &=&  1 + \frac{ \epsilon}{ 2(X_{m_1}^2 - X_{m_2}^2)} +
\sqrt{\frac{ \tilde{\alpha} (X_{m_2}^2 - X_{m_1}^2)}{ 
X (X_{m_2}^2 + X_{m_1}^2)}} + O(\epsilon^2, \tilde{\alpha})\\
t_3 &=& 1 + \frac{ \epsilon}{ 2(X_{m_1}^2 - X_{m_2}^2)} -
\sqrt{\frac{ \tilde{\alpha} (X_{m_2}^2 - X_{m_1}^2)}{ 
X (X_{m_2}^2 + X_{m_1}^2)} }+ O(\epsilon^2, \tilde{\alpha})\\
\nonumber
\end{eqnarray}

(In the above equations we assumed that $X_{m_1}^2 > X_{m_2}^2$
and $\tilde{\alpha} <0$,
which is the case in the numerical study of section 8.
Similar equations can be derived when $X_{m_1}^2 < X_{m_2}^2$). 

The solution $t=t_1$ is thus the most relevant for the study of 
the behaviour of $u$ in the flatness limit, $\epsilon \to 0$.
One finds, 
\begin{equation}
u(t_1) \sim \frac{X_{m_2}^2 - X_{m_1}^2}{2 \tilde{\alpha}} + 
\epsilon \frac{X_{m_1}^2}{2 \tilde{\alpha} (X_{m_1}^2 - X_{m_2}^2)} + 
O(\epsilon^2) \label{limitu2}
\end{equation}

which is enough to show that, as far as $\tilde{\alpha}$ is not vanishing
(which is necessarily the case when $\tan \beta$ is close to $1$, 
see Eq.(\ref{alpha})),
$u$ is not singular and 
$\zeta$ remains of order $\tilde{\alpha}$. The limiting behaviour of $u$
in Eqs.(\ref{limitu1}, \ref{limitu2}) will allow also to give an explanation 
for
the fact that the instability of the vev's in the vicinity of the D-flatness 
is much less pronounced at one-loop order than at tree-level, as was 
noted in \cite{gamb}. 
We will come back to this point again in the numerical 
discussion of section 9.\\

Let us now state some further properties which allow to determine the
relative magnitudes of $\sqrt{t_{\pm}}$ as compared to those of
$T_{\pm}$ and to $\frac{m_t}{m_b}$. $\zeta (t)$ has the nice property
of being an increasing (resp. decreasing) function of $t$ when 
$\tilde{\alpha} > 0$ (resp. $\tilde{\alpha} < 0$).

An immediate consequence of this and of the generic behaviour
of $t_{\pm}(\zeta)$ (see Fig. 1), is that the maximal
value of $t_{+}$ is $t^{max}_{+}= t_{+}(\zeta(\frac{m_t}{m_b}))$. 
By the same token,
one has $ t_{+}^{min} = t_{+}(\zeta(T_{+})), t_{-}^{max}= t_{-}(\zeta(T_{-}))$ 
and
$t_{-}^{min}= t_{-}(\zeta(t \to 1))$. 
From this one can show that $\sqrt{t_{+}^{min}}$
can never get greater than $\tan \beta_{+}$ whatever the sign of 
$\tilde{\alpha}$ is. This has the important consequence that the $t_{\pm}$
bounds alone cannot exclude the region $\tan \beta >1$, while the $T_{+}$
bound can do so, {\sl viz.} when $T_{+} > \frac{m_t}{m_b}$ (with 
$\tilde{\alpha} < 0$) or $T_{+} <  \frac{m_t}{m_b}$  
(with $\tilde{\alpha} > 0$). In contrast, one can determine at least sufficient
conditions, for instance in the region 
$X_{m_1}^2 + X_{m_2}^2 + 2 X_{m_3}^2 <<1$, for which 
$\sqrt{t_{-}^{min}} < T_{-}$, thus implying that the $\tan \beta <1$ region is 
fully 
excluded in this case. Furthermore, these conditions can be turned into 
exclusion domains in the $(Y_t, Y_b)$ plane.\\

{\bf The model-independent $\tan \beta$ bounds:}

Let us summarize here the new lower and upper bounds we found for $\tan \beta$.
We rewrite them in a way which
explicates the relevance of the sign of $\tilde{\alpha}$ and 
assume without loss of generality $X_{m_3}^2$ to be negative
(see appendix A for a related discussion): 

{\bf a)} $\tilde{\alpha} \leq 0$
\begin{eqnarray}
&\mbox{\, if \,} \tan \beta > 1 \;\;\; \mbox{then} \;\;\;T_{+} \leq \tan \beta \leq
\frac{m_t}{m_b} \;\;\;\mbox{{\bf and}} \;\;\;\tan\beta \geq \sqrt{t_{+}} \\
& \nonumber 
\end{eqnarray}
\begin{eqnarray}
&\!\!\!\!\!\!\!\!\!\!\!\!\!\!\!\!\!\!\!\!\mbox{\, if \,} \tan \beta < 1& \;\;\; 
\mbox{then} \;\;\; T_{-} \leq \tan\beta <1 \;\;\;\mbox{{\bf and}} \;\;\;
\tan\beta \leq \sqrt{t_{-}}
 \\
\nonumber
\end{eqnarray}

{\bf b)} $\tilde{\alpha} \ge 0$

\begin{eqnarray}
& \frac{m_t}{m_b}\leq \tan 
\beta\leq T_{+}& \mbox{{\bf and}} \;\;\;\tan\beta \;\;\;\geq
\sqrt{t_{+}} \\
\nonumber
\end{eqnarray} 
 
where 
\begin{equation}
T_{\pm} = \frac{ -X_{m_1}^2 - X_{m_2}^2 \mp \sqrt{(X_{m_1}^2 + X_{m_2}^2)^2 -
4  X_{m_3}^4 }}{2 X_{m_3}^2} \nonumber \\
\end{equation}

and

\begin{equation}
t_{\pm}= \frac{ \tilde{\alpha}^2 \frac{v_1 v_2}{X_{m_3}^2} - X  \mp
\sqrt{( X - \tilde{\alpha}^2 \frac{v_1 v_2}{X_{m_3}^2})^2 + \tilde{\alpha}^2-
X^2 }}{ \tilde{\alpha} - X }
\label{newcond2pf}
\end{equation}

with $0 < T_{-} \leq 1 \leq T_{+}$ when the potential is bounded from below,
and $ 0 < t_{-} \leq 1 \leq t_{+} $ in all cases ( see Fig.1). Here we also 

We also note that  the slight dissimilarity
between  $\tan \beta < 1$ and $\tan \beta > 1$ in {\bf a)} should not be
seen as contradicting the invariance of the Lagrangian 
(and the effective action) under the substitutions $ top \leftrightarrow bottom,
H_2 \leftrightarrow H_1$. Indeed this invariance is easily seen to hold 
in Eqs.(\ref{vloop}, \ref{EWSBnew} --\ref{eqv1v2}), but the point is that
$\tan \beta$ and $\cot \beta$ are not simultaneous solutions of equation
(\ref{eqtbeta}), in contrast with the tree-level case where the two $\tan \beta$
solutions are inverse of each other, Eq.(\ref{soltbeta}).\\

{\bf Excluding $\tan \beta=1$ :} We close this section by remarking
that $\tan \beta=1$ should be excluded as corresponding to a physically 
unsatisfactory situation. Actually the $t_{\pm}$ bounds exclude generically
a region around this value. Still, we want to stress here that 
$\tan \beta =1$ should be excluded on more general grounds. 
When $v_1=v_2 \neq 0$,
Eq.(\ref{EWSBnew}) implies $X_{m_1}^2 + X_{m_2}^2 + 2 X_{m_3}^2 = 0$. The 
latter implies in turn that the effective potential Eq.(\ref{vloop}) becomes
flat ($V=0$, barring of course a cosmological constant)
everywhere along the direction $v_1 = v_2$. 
The same is true at tree-level as was already remarked in section 3.
However, while at tree-level any point along the $\tan \beta =1$ was indeed
a stationary point, this degeneracy is now lifted as can be easily
seen from Eqs.(\ref{EWSBnew}), and the only EWSB stationary point 
along this direction is given by
\begin{equation}
v_1^2= v_2^2 = \frac{ X_{m_2}^2 + X_{m_3}^2}{\tilde{\alpha}} =
              -\frac{ X_{m_1}^2 + X_{m_3}^2}{\tilde{\alpha}}
\label{tbeta1}
\end{equation}

The point is that this stationary solution is not stable and cannot be
a minimum. One can see this easily in the particular case at study, from
the fact that Eq.(\ref{newinv3}) is not satisfied when $v_1=v_2$.
[Note that $\tilde{\alpha} \neq 0$ when $v_1=v_2$ since $\frac{m_t}{m_b} >>1$.] 
However we stress that there is a stronger reason which might
suggest that the result continues to hold at higher orders of perturbation 
theory: 
given our approximate treatment of the logs as described in section 2,
we find that to one-loop order 
the degeneracy is lifted but the flatness is maintained. 
The only way to go along
with this is that the shape of the potential be twisted around the point
given by Eq.(\ref{tbeta1}). This means that this solution is a saddle-point
rather than a minimum and should thus be excluded.

\newpage
\begin{center}
\section{Meaning of the new bounds}
\end{center}
\setcounter{equation}{0}
In this section we comment on the relation between the bounds we derived and
some of the features in the existing literature. To start with, it should be 
clear that the $\tan \beta$ bounds {\bf a)} and {\bf b)} reflect but partially
the constraints coming from Eqs.(\ref{EWSBnew}). This is why $X_{m_1}^2, 
X_{m_2}^2$ and $X_{m_3}^2$ are considered as three free parameters at this 
level. This means that the $\tan \beta$ bounds delimit the regions for which
a stationary point is an actual (local) minimum
with non-vanishing vev's. On the other hand, the $\tan \beta$ bounds contain
the full information coming from the $2^{nd}$ order derivatives at the one-loop
level Eqs.(\ref{newinv3}, \ref{newcond1}),  
(keeping in mind that we absorbed the logarithmic terms in the tree-level
part). They thus imply that the one-loop corrected Higgs mass-sum-rules, which
have been extensively studied in the literature 
\cite{ellis1,ellis2,comp, habertalk}, 
are such that the
requirement of positivity of the squared Higgs masses leads in general to 
non trivial constraints. From this point of view, it is to be noted that
if the one-loop minimum remains sufficiently close to its tree-level
counterpart, then it is likely that these positivity conditions be automatic
as is precisely the case at the tree-level. This was for instance the case in
\cite{ellis1} where the D-terms in the stop mass matrix where neglected and 
the renormalization scale
chosen such that the stationarity condition be fulfilled at the tree-level.
The leading top-stop effect in the one-loop correction to the lightest
CP-even Higgs squared mass turned out then to be always positive.
More generally however, keeping those D-terms (which lead to the 
$\tilde{\alpha}$ contributions) one
should rely on the one-loop stationarity condition, and as can be seen for
instance from ref.\cite{ellis2} the positivity is no more automatic. 
Conditions {\bf a)} and {\bf b)} are
the analytic constraints (within our approximation) which ensure this
 positivity.\\

Let us now compare briefly  some of the well-known constraints, namely
$1 \leq \tan \beta \leq \frac{m_t}{m_b}$, and our condition {\bf a)}.
We recall first that $1 \leq \tan \beta \leq \frac{m_t}{m_b}$ is a 
model-dependent result in the context of SUGRA-GUT \cite{giudice}
obtained through the running of $m_1^2$ and $m_2^2$ 
as dictated by the theory, in Eq.(\ref{solv1v2}).
Actually one can easily see from Eq.(3.17) and what precedes it, that
in order to exclude $\tan \beta <1$ one needs either $m_2^2 <0$
or $m_1^2 > m_2^2$. The first of these inequalities is a sufficient but
not necessary condition for electroweak symmetry breaking 
to take place, so that one also
needs to consider the condition $m_1^2 > m_2^2$ in order to exclude
$\tan \beta <1$ in all cases. This is where, besides the running of $m_1^2$ and $m_2^2$, 
enters the universality assumption $m_1^2(M_{GUT}) = m_2^2(M_{GUT})$
leading to  $m_1^2 > m_2^2$ at the electroweak scale.
In our case we assume none of the above model-dependent features. Yet, 
though fully model-independent, our condition
$1 \leq Max(T_{+}, \sqrt{t_{+}}) \leq \tan \beta \leq \frac{m_t}{m_b}$ appears
to be stricter than the previous one since it excludes a calculable
domain around $\tan \beta=1$. There is of course no contradiction here; it only means 
that using the tree-level (actually tree-level-RGE-improved) 
minimization conditions together with some model-dependent assumptions
is weaker than just using the model-independent one-loop minimization 
conditions. Of course, one can still make the same
model-dependent assumptions on top of the one-loop minimization conditions
we derived and get even stronger bounds.  
Also our bounds should be contrasted with the more qualitative ones,
derived in the literature from the requirement of perturbativity of the 
Higgs-top or Higgs-bottom Yukawa couplings, \cite{tbetaperturb}. 
It is clear that
the requirement of perturbativity is more of practical than physical
relevance, while our constraints are directly related to the physical
requirement of symmetry breaking.

\begin{center}
\section{Comments on necessity and sufficiency}
\end{center}
\setcounter{equation}{0}
The constraints and bounds we established in section 4 are necessary for
the existence of local minima which break $SU(2)_L \times U(1)_Y$ to
$U(1)_{e.m}$. If there exists more than one such minimum, the true minimum will 
be of course the lowest, or at least a sufficiently metastable one. 
However, when it comes to assess the sufficiency
of our conditions several other questions should be asked.
First, does the symmetric vacuum become unstable as a consequence of the 
existence
of an electroweak vacuum as was the case at tree-level (see Eqs.(3.14, 3.15))?
At vanishing $H_1$ and $H_2$, the invariants of the matrix of second order 
derivatives of the one-loop effective potential
with respect to the eight components of $H_1$ and $H_2$, are all equal to
$${\cal I}=X_{m_1}^2 X_{m_2}^2 - X_{m_3}^2$$ which is a direct generalization of
the tree-level case, see Eqs.(3.14, 3.15). Substituting for $X_{m_1}^2$ and
$X_{m_2}^2$ through Eq.(\ref{EWSBnew}) one finds

\begin{equation}  
{\cal I}= - \tilde{\alpha}^2 v_1^2 v_2^2 + (v_2^2 - v_1^2) [
\tilde{\alpha} ( v_1^2 + v_2^2) - ( v_2^2 - v_1^2) X] (X - 
\frac{X_{m_3}^2}{v_1 v_2})
\end{equation}
Upon use of Eqs.(\ref{newinv3}, \ref{newcond1})
one then shows easily that

\begin{equation}
{\cal I} \leq \tilde{\alpha}^2 v_1^2 v_2^2 
(2 \frac{v_1 v_2}{X_{m_3}^2} X - 3) \leq 0
\label{unstabioneloop}
\end{equation}

Thus the one-loop conditions for the existence of EWSB local minima are 
{\sl sufficient} to make the gauge symmetric vacuum unstable. 
It should be clear that this is
not a self-evident statement. There would be nothing wrong with a
local minimum at the gauge symmetric point, but then one would have had to
impose extra constraints for this minimum not to be the global minimum.

Up to now all the discussion was lead in the electrically neutral
$H_1, H_2$ directions defined  by 
Eq.(\ref{neutrdir}). Charge breaking directions like
\begin{eqnarray} 
H_1= \left( \begin{array}{c} 
               0 \\
               H_1^{-} \\
               \end{array} \right) \hspace{2cm}
H_2= \left( \begin{array}{c} 
               0 \\
               H_2^0 \\
               \end{array} \right) \label{chargdir} \\
\nonumber
\end{eqnarray}

are usually assumed to be devoid of minima.  
Indeed, such minima
do not occur at tree-level or ``tree-level improved'' approximation in the
MSSM. They do however occur in a general two-higgs-doublet model,
and imply in this case some constraints on the tree-level Higgs couplings
\cite{sher}. Furthermore, 
knowing of no general proof guaranteeing the absence of such minima 
at higher loop orders
even in the MSSM, we have to study the one-loop case explicitly.
It is instructive to show briefly what the typical pattern of first
and second order derivatives looks like and how it changes when one
goes from tree-level to one-loop order. We will take for illustration 
the direction defined in Eq.(\ref{chargdir}), and assume furthermore
$ Im(H_1^{-}) = Im (H_2^0) = 0$ for simplicity. We denote the vev's
$<Re(H_1^{-})>$ and $< Re(H_2^0)>$ respectively by
$c_1$ and $c_2$. Demanding that $c_1$ and $c_2$ be simultaneously non-vanishing
requires $X_{m_3}^2 =0$ and leads to the following stationarity conditions

\begin{eqnarray}
2 ( X - \tilde{\alpha} ) c_2^2 + ( \tilde{\beta} - 2 X) c_1^2 + 2 X_{m_2}^2&=&0
 \label{dcharge1}\\
2 ( X + \tilde{\alpha} ) c_1^2 + ( \tilde{\beta} - 2 X) c_2^2 + 2 X_{m_1}^2&=&0
\label{dcharge2}\\
\nonumber
\end{eqnarray} 

The above equations are also valid at tree-level, the replacement of 
$\tilde{\alpha}, \tilde{\beta}, X, X_{m_i}^2$ by their  tree-level counterparts
being understood. Since $ \tilde{\beta} \pm 2 \tilde{\alpha}$ is always {\sl
perturbatively} positive valued,
summing Eqs.(\ref{dcharge1}, \ref{dcharge2}) and taking into account the fact
that $c_1^2 $ and $ c_2^2$ are positive, implies readily that 
$X_{m_1}^2 + X_{m_2}^2 <0$. The last inequality together with $X_{m_3}^2 =0$
violates Eq.(\ref{ufbloop}) and leads to an unbounded from below effective
potential. If one takes the unboundedness from below of the MSSM as a disaster
then it should be concluded that charge breaking minimum in the Higgs sector
cannot occur in realistic situations. Indeed in sections 3 and 4 we assumed
boundedness from below all the way. From this point of view, combining 
Eq.(\ref{unstabioneloop}) and the absence of charge breaking just illustrated
one concludes that {\sl the necessary EWSB conditions} derived in 
those 
sections {\sl are also sufficient, at least in the Higgs fields directions}.\\

From a more general point of view, though, one should keep open the possibility
that the full theory can still be well behaved even in the case where the MSSM 
has an unbounded from below potential. After all, the MSSM is only a low-energy
effective theory while the behaviour of the effective potential at very large
field values becomes sensitive to the heavy ($\sim M_{GUT}$) degrees of 
freedom whose interactions might cure the apparent unboundedness from below.
In this case one should study further the second order derivatives to question
the occurrence of charge breaking minima. An important difference
shows up between the tree-level and one-loop cases, of which we give here
just a glimpse. In both cases one finds only three independent invariants
of the second order derivative matrix. At tree-level, two of these invariants
have the following form:
\begin{eqnarray}
{\cal C}_1^{tree}&=& m_1^2 + m_2^2 \\
{\cal C}_2^{tree}&=& - \frac{g_1^2 + g_2^2}{g_2^2}(m_1^2 + m_2^2) \\
\nonumber
\end{eqnarray}
and are thus always of opposite signs irrespectively of the question of
unboundedness from below. Thus at tree-level there is no way of having
a (even local) charge breaking minimum. [Note also that 
${\cal C}_1^{tree}={\cal C}_2^{tree}=0$ would lead through 
Eqs.(\ref{dcharge1}, \ref{dcharge2}) to zero charge breaking vev's.]\\
The situation is more dangerous at one-loop order. The three invariants
read in this case

\begin{eqnarray}
{\cal C}_1^{1-loop}&=& ({\cal R}_1 + {\cal R}_2) \frac{\tilde{\beta}}
{4 \tilde{\alpha}^2 + \tilde{\beta}^2 - 4 \tilde{\beta} X} \\
{\cal C}_2^{1-loop}&=& - \frac{4 {\cal R}_1 {\cal R}_2}{ 
4 \tilde{\alpha}^2 + \tilde{\beta}^2 - 4 \tilde{\beta} X} \\
{\cal C}_3^{1-loop}&=& (\tilde{\alpha} - X) (2 \tilde{\alpha} - \tilde{\beta} +
4 X) X_{m_1}^2 + (\tilde{\alpha} + X) (2 \tilde{\alpha} + \tilde{\beta} -
4 X) X_{m_2}^2 \\
\nonumber
\end{eqnarray}         

where 
\begin{eqnarray}
{\cal R}_1&=&2 ( \tilde{\alpha} - X) X_{m_1}^2 + 
(\tilde{\beta} - 2 X) X_{m_2}^2
\\
{\cal R}_2&=& (\tilde{\beta} - 2 X) X_{m_1}^2 - 2 ( \tilde{\alpha} + X)\\
\nonumber
\end{eqnarray} 

and give back, of course, the tree-level results when $\tilde{\alpha} \to 0, 
\tilde{\beta} \to g_2^2/2, X \to (g_1^2 + g_2^2)/8$ and $X_{m_i}^2 \to m_i^2$.
However one can find now a region, in the perturbative regime, where
${\cal C}_1^{1-loop} \geq 0 $, ${\cal C}_2^{1-loop} \geq 0 $, 
${\cal C}_3^{1-loop} \geq 0 $ are simultaneously verified and consistent
with $X_{m_1}^2 + X_{m_2}^2 \leq 0$ which is required by Eqs.(\ref{dcharge1},
\ref{dcharge2}). This happens if $|\tilde{\alpha} - X|, \tilde{\alpha} + X
< \tilde{\beta} -2 X$. The point is that the perturbative realization of 
these last inequalities depends on the relative magnitudes of $g_1^2$ and
$g_2^2$. For instance they are not satisfied at the unification scale
where one has typically $g_2^2 = (5/3) g_1^2$, but they are satisfied
at the electroweak scale where $g_2^2 \sim 3.3 g_1^2$. If anything,
this means that, provided the apparent unboundedness from below is tolerated
(see also the related comment at the end of section 8),
local charge breaking minima can occur in the Higgs directions of the MSSM
at the one-loop level! One would then have to require such minima to remain
higher than the electroweak one and to be sufficiently short lived.
We do not dwell further on these aspects here.\\

Before closing this section we should note that one has in principle to study 
the stationarity and stability of the electroweak minimum in all squark/slepton
directions as well. These aspects will be considered elsewhere. We will be 
contented
here just with the remark that at tree-level the curvature of the effective 
potential
is, on one hand, proportional to the squared squark (or slepton) masses
in the respective squark (or slepton) directions at the zero fields point,
and on the other, that this curvature remains positive in these directions also at the
electroweak point, for realistic values of $m_1, m_2$ and squark/slepton masses.
This does not mean, though, that no charge or color breaking minima can occur,
since one can still be driven around a hump into a minimum in a charged or colored
direction.

\begin{center}
\section{Gauge-fixing dependence}
\end{center}
\setcounter{equation}{0}
In this section, we make a short digression and discuss briefly the issue
of gauge-fixing dependence of the constraints established so far.
It is well-known that the effective potential $V_{eff}$ is generally a 
gauge-fixing dependent object, even though it is formed of gauge 
invariant operators. However, the values of $V_{eff}$ at {\sl stationary} 
points,
corresponding to average energy densities of ``vacuum'' states,
are themselves gauge-fixing independent 
at least for a given class of gauges \cite{xsindep}. 
[This result holds perturbatively in the loop-wise expansion]. 
It follows that the condition $\frac{\partial V_{eff}}{\partial \phi} = 0 $ 
selects 
physical stationary points, even though $\frac{\partial V_{eff}}{\partial \phi}$ 
as well as the field values at which it vanishes (including the vev's), are
generically gauge-fixing dependent. \\
We want to point out hereafter that, under fairly general assumptions about 
$V_{eff}$,  a (local) minimum or maximum, remains
so for any value of the gauge-fixing parameter, even if the magnitude of
$\frac{\partial^2 V_{eff}}{\partial \phi^2}$ is itself gauge-fixing dependent.\\
Let us assume that $V_{eff}( \phi, \xi)$ is a smooth continuous 
function of $\xi$,
where $\xi$ is a generic gauge parameter, and that $V_{eff}$ is bounded
from below. As far as boundedness from below is concerned, one can think of 
this condition as requiring something like Eq.(\ref{ufbloop}) simultaneously 
for 
any value of the $\xi$ parameter\footnote{This avoids questions related to 
whether the necessary and sufficient boundedness from below conditions
are gauge-fixing dependent or not.}. We can now show that a (local) minimum
for a given value $\xi_1$ of the gauge parameter, cannot transform into a 
(local) maximum or a saddle-point for another value $\xi_2$. 
Indeed, since $V_{eff}$ is $\xi$-independent at every stationary point 
\cite{xsindep},
the only way of changing a (local) minimum at a point M,
 when $\xi_1$ goes to $\xi_2$, is to
create a new stationary point M' which corresponds to a {\sl lower} value of
$V_{eff}$, and which lies between M and the next  
nearby stationary point.
However, this new stationary point (which exists for $\xi_2$ and not for 
$\xi_1$) must be degenerate with one of the already pre-existing stationary 
points
at the value $\xi_1$, otherwise it would contradict the 
$\xi$-independence of $V_{eff}$ at stationary points. But even if such
a degeneracy occurs, $V_{eff}$ being
a smooth continuous function of $\xi$ there will always be
a value of $\xi$ between $\xi_1$ and $\xi_2$ for which a new stationary point
appears and is not (yet) degenerate with any of the previously existing ones,
thus contradicting the aforementioned $\xi$-independence. This completes the 
``proof'' that if  $\frac{\partial^2 V_{eff}}{\partial \phi^2} > 0$ 
at stationary points for a given
value of $\xi$, then it is fulfilled for any value of $\xi$. A similar 
reasoning applies when there are more than one scalar field, and also in the 
case of (local) maxima or saddle-points.
We illustrate in Fig.2 a typical behaviour of the effective potential
under a gauge-fixing change. 

The above discussion shows that the perturbative determination of the 
conditions for the occurrence of (local) minima is physically meaningful,
despite the $\xi$-dependence of the quantities we compute 
({\sl i.e.} first and second order derivatives). 
Nonetheless,the fact that the vev's
are $\xi-dependent$ requires some precaution when relating them to
the (physical) electroweak scale. For instance, when $M_Z^2$ is equated
to $\sim (g_1^2 + g_2^2)(v_1^2 + v_2^2)$ to obtain Eqs.(\ref{EWSBcond})
as the condition of symmetry breaking at the correct electroweak scale,
one should keep in mind that this is but a gauge-fixing dependent approximation.
The physical mass $M_Z$ is the mass pole which is not given 
(beyond the tree-level) by the quantity  $(g_1^2 + g_2^2)(v_1^2 + v_2^2)$,
the latter being $\xi$-dependent, \cite{xsindep}.


\begin{center}
\section{Numerical illustration}
\end{center}
\setcounter{equation}{0}
The one-loop EWSB equations and constraints established in 
section 4 should be very helpful in the implementation of any radiative
symmetry breaking study in the MSSM. They are fully model independent, 
and must be adapted in each model which triggers electroweak symmetry 
breaking. In this section we give a very first 
illustrative example in a SUGRA-GUT context. Since the aim is just to
compare the tree-level and one-loop non-logarithmic contributions, we
will take up here, as discussed in section 2,
the simple (and somewhat old fashioned) approximation
of the renormalization group improved tree-level, without taking into 
account
supersymmetric threshold effects which would not however modify the 
qualitative picture.\\ 
We show how the various bounds and constraints presented in section 4 
conspire to give, as in the the tree-level case, 
only one local minimum of the potential.
This should not come as a surprise. At the tree-level, the equation 
for $ tan\beta$ is quadratic
and gives at most two real solutions. Actually, we fall in the 
situation depicted 
in section 3, where one solution gives $0 <\tan \beta <1 $. Together 
with ${m_2}^2 < 0$, this solution is thus rejected for $u\equiv 
v_1 v_2$  becomes negative. 
This situation is common in SUGRA-GUT motivated analyses. 
Universality for the soft scalar mass terms combined with the relative 
rapid decrease with the scale of ${m_2}^2$ in comparison with ${m_1}^2$ 
( due to the contribution of $Y_t$ in its renormalization group equation) 
drives ${m_2}^2$ negative while ${m_1}^2$ remains positive. By tuning 
initial values at GUT scale, we can arrange to have
$ {m_1}^2{m_2}^2 -{m_3}^2 <0$
near electroweak scale. This is the signal that the gauge-invariant 
vacuum is unstable, leading to the electroweak symmetry breaking ( we 
discard here 
the possibility of CCB minima outside the higgs direction). 
Now adding to the tree-level potential one-loop non-logarithmic contributions, 
the equation for $\tan \beta$ Eq. (4.13) becomes
quartic. So we can  potentially have up to four different and real
solutions in this case. However all of them do not correspond
to local minima, and the bounds established in section 4 are precisely 
here to reject all of them but one which is near the local 
minimum obtained 
at the tree-level. \\
This is not the only point of the discussion. $\tilde{\alpha}$  also 
contributes to moderate the behaviour of $v_1$ and $v_2$
as compared to the well-known tree-level wild behaviour near the scale where 
the potential becomes unbounded from below. We will also show
how this comes about. \\ 
Assuming universality at some unification scale $\tilde {\alpha}, X, $ and
the $X_{m_i}^2$'s can be expressed as follows, see Eqs.(4.2-4.8) 

\begin{eqnarray}
\tilde{\alpha} &=& \frac{12}{5} \kappa g_{0}^{2}({Y_t^0}^2- {Y_b^0}^2) \\
X &=& \frac{{{\overline{g}}_1}^2(\mu_R)+{{\overline{g}}_2}^2(\mu_R)}{8} 
+ \kappa g_{0}^2({\frac{367}{100}}-{\frac{12}{5}} ({Y_t^0}^2+ {Y_b^0}^2))   \\
X_{m_1}^2 &=& {{\overline{m}}_{1}}^{2}(\mu_R) + \kappa
[{\frac{72}{15}} g_{0}^{2} m_{1/2}^2 +
{\frac{18}{15}}( m_0^2 - 3 \mu_0^2) \nonumber\\
&& +12 (\mu_0^2 {Y_t^0}^2+{Y_b^0}^2 ({A_b^0}^2   + 2 m_0^2) )] \\
X_{m_2}^2 &=& {{\overline{m}}_{2}}^{2}(\mu_R)+ \kappa
[{\frac{72}{15}} g_{0}^{2} m_{1/2}^2 +
{\frac{18}{15}}( m_0^2 - 3 \mu_0^2) \nonumber\\
&&+ 12 (\mu_0^2 {Y_b^0}^2 +{Y_t^0}^2 ({A_t^0}^2   + 2 m_0^2) ) ] \\
X_{m_3}^2&=& {{\overline{m}}_{3}}^{2}(\mu_R)+ \kappa \mu_0 
({\frac{18}{15}}( B_0 + 4 m_{1/2}) - 12 (A_t^0 {Y_t^0}^2+ A_b^0 {Y_b^0}^2)) \\
\nonumber
\end{eqnarray}
We run the tree-level part of the above equations, using the one-loop
RGE solutions for small $\tan\beta$, and keep their one-loop parts
fixed at their initial values. Here $g_{0}$ is the unified value of the 
gauge couplings (taking a grand unified normalization for the $U(1)_{Y}$ 
coupling, 
$g_{0}^2={g_{2}^0}^2=5/3 \ {g_{1}^0}^2 $). 
We take for illustration, neglecting the contribution of $Y_b^0$, the 
following values of universal soft masses
and other parameters which correspond to a no-scale scenario.

\begin{eqnarray}
&& A_0=B_0=m_0=0, \ \ \ \ m_{\frac{1}{2}}= 50 \; \mbox{GeV,} \ \ \ \ 
{\frac{{g_0}^2}{4 \pi}}= {\frac{1}{24.5}}  \\
&& \mu_0= -105 \; \mbox{GeV,} \ \ \ \ {\frac{(Y_t^0)^2}{16 \pi^2}}=0.01, \ 
\ \ \ Y_b^0\sim 0, \ \ \ \  M_{GUT}=2\times 10^{16} \ \mbox{GeV}  \\
\nonumber
\end{eqnarray}
In this 
illustration, because $\tilde{\alpha}$ is purely a one-loop effect, it will not 
vary with the renormalization scale $\mu_R$. 
The initial values above give a negative $\tilde{\alpha}$,
$\tilde{\alpha}\sim-4.6 \times 10^{-3}$. Near the electro-weak scale, 
where we will focus our attention in 
the following,
we also find   
$X_{m_3}^2$ to be negative, giving thus positive $T_{+}$ and $T_{-}$ 
(sign opposite to the one of 
$X_{m_3}^2$).\\
Varying the scale $\mu_R$ around the electro-weak scale, we then solve 
the quartic 
equation in $\tan \beta$ (4.13).
It should be clear that all the solutions (as well as the corresponding vev's)
will vary 
with  $\mu_R$. In our illustration, all of them are real, and for convenience   
we may label each one by comparing 
with the limiting case $\tilde{\alpha}=0$ where the solutions 
are just $T_{+}$, $T_{-}$, 1 and -1. But one must keep in mind that, 
as soon as $\tilde{\alpha}{\not =}0$, $T_{+}$ and $T_{-}$ are no 
more solutions, but rather bounds.\\ 
All the solutions do not correspond to local minima. In fact, 
we can safely discard the solution which is 
near -1, because being negative, it contradicts constraint (4.19).  
The solution near $T_{-}$ is rejected 
for the reason that it is included in $[0,T_{-}]$ and so contradicts 
constraint (4.35). We must also reject the solution near 1, because it 
is included in 
$[\sqrt{t_{-}}, \sqrt{t_{+}}]$, and does not verify constraints (4.34-4.35). 
We call this last solution $t_1$, for it coincides 
with Eq.(4.30) in the flat limit. Fig.3 shows its sensitivity to 
the renormalization scale and displays its theoretical bounds.\\
Note that 
to reject all those solutions, we have fully used constraints 
(4.19,4.34,4.35) which contain all the qualitative informations on 
the second stationarity 
equation (4.14). They also give information on the stability
of the possible extrema. If we had used only the two stationarity 
equations, we could have already rejected the first 
two solutions since the corresponding $u(\equiv v_1 v_2)$  
would have had a sign opposite to that of $\tan \beta$. 
The third solution, $ t_1$, would not be correctly rejected this way,
being indeed an extremum of 
the potential but not a local minimum.\\
Finally, there is only one solution that will 
satisfy all constraints.   
This is the one near $T_{+}$ which we call $t_2$ for it gives 
Eq.(4.31) in the flat limit. It is such that u  and $t_2$ have the same sign, 
and so we can be sure that it indeed corresponds to a local minimum of the 
potential. We are thus left, as in the tree-level case, with only one correct 
electro-weak minimum. \\
Fig.4 displays the variation with 
respect to $\mu_R$ of $t_2$ and its theoretical 
bounds. For comparison, we also give in the same figure  
the variation of the tree-level solution $\tan \beta_0$. \\
In the bounded from below region, that is 
above the scale $\mu_R^{UFB}\sim 52 \ \mbox{GeV } $
( defined by $X_{m_1}^2+X_{m_2}^2+2 
X_{m_3}^2|_{\mu_R^{UFB}}=0$), $t_2$ is just above 
the 
bound $T_{+}$ and follows it as the scale decreases. It is also far above 
$\sqrt{t_{+}}$ which is very close 1. Generically, we have this hierarchy
in SUGRA-motivated analysis: $\sqrt{t_{+}}$
is never greater than $T_+$, except in a small region above $\mu_R^{UFB}$. 
Note also that $\tan \beta_0$ is below 
$T_{+}$ and so does not verify the constraints (4.34). In fact, there is no 
reason that it should. The only lower bound we may give at the tree-level 
is  $\tan \beta_{0}> 1$ for ${m_{2}}^{2}<0$. 
We just want to stress here that the new stability constraints $T_+$ 
can be quite significant and forbid some 
values 
of $\tan \beta$ including large ones taken by $\tan \beta_0$.\\ 
$T_{+}$ and $T_{-}$ 
terminate at $\mu_R^{UFB}$ and take at this scale the common value 1, while
$t_2$ ( and so does $t_1$, see Fig.3) continues to exist and still
makes sense at 
and below $\mu_R^{UFB}$. This situation is quite typical of 
the one-loop non-logarithmic contributions.
On the contrary, the tree-level solution $\tan \beta_0$ terminates when the 
unboundedness from below scale $(\mu_R^{UFB})_{tree} \sim 54 \ \mbox{GeV } $
( defined by $m_1^2+m_2^2+2 m_3^2|_{(\mu_R^{UFB})_{tree}}= 0$)
is reached.\\
Finally, in the UFB region where $T_{+}$ and $T_{-}$ no more exist,
the bounds 
$\sqrt{t_{+}}$  and $\sqrt{t_{-}}$ take over from them.
As long as $ t_2$ remains above $\sqrt{t_{+}}$ in Fig.4, the corresponding 
stationary
point is indeed a local minimum with finite vev. There is a critical scale 
where $ t_2$ meets $\sqrt{t_{+}}$. If we look back in Fig.3, we see 
that this phenomenon
also appears for $t_1$, at the same scale and 
with the same value $ t_1=t_2 \sim 1.12$. In this case the extremum could
become 
a local minimum of the potential and we may wonder if, below the 
common critical scale, $t_1$ could take over on $t_2$. In fact, this does not
happen: both $t_1$ and $t_2$ stop to be real valued, and this is 
the reason why, in Fig.3-4, the curves $t_1$ and $t_2$ suddenly stop. This 
behaviour of the solutions in $tan \beta$ of Eq. (4.13) is typical of 
quartic equations, and always takes place in the UFB region.\\
Fig.5 gives the variation with 
respect to $\mu_R$ of the vev $v_1$ and $v_2$ associated with the tree-level
solution $\tan \beta_0$ and with the one-loop solution $t_2$. \\
A few comments are in order so as to compare the tree-level approximation with
the one-loop non-logarithmic one.
The curves obtained are
essentially similar when the potential is bounded from below, but behave
quite differently in each approximation near the UFB region. This point is of 
course linked to the 
preceding discussion on the $tan \beta$ solutions. 
At the tree-level,  $v_1$ and $v_2$ blow up in the vicinity 
of $(\mu_R^{UFB})_{tree}$ which 
is reached
only asymptotically and never crossed. 
This well-known wild behaviour depends on the values 
taken by the mass parameters that enter the tree-level potential near this 
scale. A quick look at the extremal equations Eq. (3.3,3.4) 
shows that u remains finite if $\tan \beta=1$ only for 
particular values of 
the mass parameters. We must realize $ m_1^2+m_2^2+2 m_3^2= 0$ 
with the requirement that $m_1^2 = m_2^2= - m_3^2$. In a SUGRA-GUT context, 
this is obviously impossible, and even if we deviate from 
universality for the soft scalar masses, this last condition should be 
adjusted by a fine-tuning which seems rather unnatural. In our illustration,
at $(\mu_R^{UFB})_{tree}$  we have 
$ m_1^2 \sim 6457 \ \mbox{GeV }$, 
$ m_2^2 \sim -3133 \ \mbox{GeV }$, $ m_3^2 \sim -1662 \ \mbox{GeV }$ which
makes  $ m_1^2+m_2^2+2 m_3^2 \sim 0$. The tree-level potential becomes flat 
and vanishing
in the direction $v_1=v_2$ and $\tan \beta=1$ is the only solution
toward which $\tan \beta_0$ is attracted. But the condition
$ m_1^2=m_2^2=-m_3^2$ is not fulfilled, so the flat direction will be
reached only asymptotically with 
exploding $ v_1$ and $ v_2$.\\
In contrast,
Fig.5 shows that in the one-loop non-logarithmic approximation, $v_1$ and $v_2$
do not have a singular behaviour near $\mu_R^{UFB}$. We can understand this 
point by noting that $t_2$ is not attracted to 1, 
(see Eq.(4.31) and Fig.4), so u remains finite.[Note however that 
$\tan \beta=1$ is not a singular value for u, as can be seen in Eq.(4.33), 
because of the genuine effect of 
$\tilde{\alpha}$, whatever its 
magnitude.] 
Although the vev's vary rapidly near the "critical scale" (where
$t_2$ stops to exist), they still remain finite. Below this scale the
vev's become complex valued, and disappear in Fig.5.\\
Regarding the mild behaviour of the vev 
in comparison with the tree-level one, 
we can say that 
adding
non-logarithmic terms to the tree-level potential, even before including 
logarithmic ones, 
already contributes to stabilize the vev of the Higgs fields.

{\bf Remarks on unboundedness from below}
One might inquire whether it is physically relevant to consider, as we did
in the previous paragraph, the regions of the parameter space where the
effective potential is unbounded from below. Considering the MSSM 
as a low-energy approximation of a more fundamental theory which would set 
in at high energy (presumably $M_{GUT}$), one should keep in mind that
our effective potential is also ``effective'' in the sense that only
light degrees of freedom are taken into account. The fact that we consider 
regions where equation 
(\ref{ufbloop}) is violated means that we implicitly assume the heavy degrees
of freedom of the full theory to play an important role in lifting the 
effective
potential at very large values ($ \sim M_{GUT}$)of the scalar fields and 
restore a bounded from below potential.
In this case, the relevant question one can ask from a model-independent point
of view is whether local minima exist even when $V_{eff}$ is a decreasing, and
 negative valued
function at large field values, and whether any of these minima can be lower
than the value of $V_{eff}$ for fields just below the scale where the full 
theory should set in.    
 
\begin{center}
\section{Conclusion}
\end{center}
We have demonstrated in this paper, the existence of fully 
{\sl model-independent} constraints involving $\tan \beta$ and the other
free parameters of the MSSM. These constraints are derived from a careful
study of the EWSB conditions, including the finite (non-logarithmic)
one-loop corrections to the effective potential. We have shown with this
respect that stationarity and stability conditions ( which are necessary
and sufficient to ensure the existence of a local minimum) are not
automatically related beyond the tree-level approximation. We then determined
the exact analytic forms of both and deduced new bounds on $\tan \beta$.
These bounds are, generally speaking, resemblant to the existing ones,
but they turn out to be more restrictive even though
fully model-independent! They also distinguish quite naturally the
``large'' and ``low'' $\tan\beta$ regimes, restricting further the allowed
range within each of these regimes.\\
As a first step, we resorted in this paper to the approximation where the
logarithmic contributions are taken into account through the
RG-improved tree-level potential. This, we believe, contains already
many of the qualitative features of the problem at hand. The inclusion
of the logarithmic effects beyond the present approximation will be
presented elsewhere.

Finally, it is clear that the existence of model-independent constraints
is interesting both theoretically and phenomenologically. For instance we
noted that the connection between stability and stationarity, which is
lost beyond the tree-level, can be partially preserved if 
$1 < \tan \beta <\frac{m_t}{m_b}$. If this connection is a desirable theoretical
criterion then the model-independent study helps select the model assumptions
likely to satisfy it. More down to earth, given the variety of theoretical
models and the fact that recent data seem to
favour some of the MSSM parameter values that are not consistent with the 
conventional model assumptions \cite{kane}, the model-independent constraints
would help narrow down the domains that should be common features to all 
models. In particular our constraints to one-loop order are readily applicable
in the context of R-parity violation, and can be generalized to the
next to minimal MSSM.

\section{Acknowledgments}
We would like to thank Nikolai Krasnikov for asking the question which lead
to section 7. We also acknowledge useful discussions with 
Michel Capdequi-Peryranere, Abdelhak Djouadi and Jean-Lo\"{\i}c Kneur.
\newpage
\renewcommand{\thesection}{Appendix}
\begin{center}
\section{}
\end{center}
\renewcommand{\theequation}{A.\arabic{equation}}
\renewcommand{\thesection}{A:}
\begin{center}
\section{The $\tan \beta$ bounds}
\end{center}
\setcounter{equation}{0}

Starting from Eq. (\ref{eqv1v2}), 
\begin{equation}
u = \frac{1}{\tilde{\alpha} (t^2  -1)} a(t)
  \label{eqv1v2p} 
\end{equation}
with $u \equiv v_1 v_2$, $ t\equiv \tan \beta$ and
\begin{equation}
a(t)=  X_{m_3}^2(t^2  + 1) + 
(X_{m_1}^2 + X_{m_2}^2 )t
\end{equation}  
and assuming boundedness from below 
Eq (\ref{ufbloop}), we note first that the two zeros of $a(t)$, i.e.

\begin{equation}
T_{\pm}= \frac{-X_{m_1}^2 - X_{m_2}^2 \mp \sqrt{(X_{m_1}^2 + X_{m_2}^2)^2 
-4 X_{m_3}^4}}{2 X_{m_3}^2}
 \label{Tpm}
\end{equation}
have the same sign, {\sl opposite to that of $X_{m_3}^2$} . Furthermore one has
\begin{equation}
|T_{-}| < 1 < |T_{+}|
\label{Tpmprop} 
\end{equation}
the degenerate value $1$ being attained when the inequality Eq.(\ref{ufbloop})
is saturated. On the other hand, we note that
\begin{equation}
\tilde{\alpha} < 0 \Leftrightarrow |Y_t|>|Y_b| \Leftrightarrow 
|\tan \beta| <\frac{m_t}{m_b}
\label{signalpha}
\end{equation}
as easily seen from Eqs.(\ref{alpha}) and (4.9) and from the usual relations
$ m_t= \frac{|Y_t v_2|}{\sqrt{2}}, m_b = \frac{ |Y_b v_1|}{\sqrt{2}} $.
We keep here absolute values on purpose (and somewhat puristically),
 since we would like to establish clearly the freedom
in choosing the sign of $\tan \beta$, which is rather tricky in the present
context. 

In the following we will address the question of whether the necessary stability
condition $\frac{X_{m_3}^2}{v_1 v_2} \leq 0$ (Eq.(\ref{newcond1})) 
is automatically fulfilled,
and will establish in the process some model-independent analytic bounds
on the allowed values of $\tan \beta$. The properties of $T_{\pm}$ 
presented above, and the sign of $\tilde{\alpha}$ 
will be crucial for the subsequent discussion. 

Regarding Eq.(\ref{signalpha}) we should consider the two cases
\begin{equation}
|\tan\beta| \leq \frac{m_t}{m_b} \;\;\;\mbox{and}\;\;\; 
|\tan\beta| \geq \frac{m_t}{m_b}
\end{equation}
and in the former case, we study separately 
$\tan^2 \beta >1 $ and $\tan^2 \beta <1 $.\\

{\bf Case 1)} $|\tan\beta| \leq \frac{m_t}{m_b}$, ($ \tilde{\alpha} < 0$).\\

{\bf (1-a) $\tan^2 \beta >1 $ } implies $\tilde{\alpha} (t^2  -1)<0$.
It follows that in Eq.(\ref{eqv1v2p}),
$ a(t)$ and $u$ [and thus $t$]
have {\sl opposite signs}. This means that if $ X_{m_3}^2 \leq 0$ then
$$ \tan \beta \geq T_{+} \geq 1$$
is the only allowed region. Indeed, on one hand,
$$ 0\leq T_{-} \leq \tan \beta \leq T_{+}$$
would imply that $a \geq0$, {\sl i.e.} of the same sign as $\tan \beta$,
and on the other hand, if $\tan \beta \leq T_{-}$ (which means that
$\tan \beta \leq -1$, since $\tan^2 \beta >1 $ and $T_{-} \leq 1$)
then $a<0$ . Thus again $a$ and $\tan \beta$ would have the same sign,
which is forbidden.

Similarly, if $ X_{m_3}^2 \geq 0$ then 
$$ \tan \beta \leq T_{+} \leq -1$$ is the only possibility compatible
with the requirement that $a$ and $\tan \beta$ be of opposite sign.

To summarize, one has
\begin{eqnarray}
1 \leq T_{+} \leq \tan \beta \;\;\; &\mbox{if}&\;\;\; X_{m_3}^2 < 0 \nonumber \\
\tan \beta \leq T_{+} \leq -1 \;\;\; &\mbox{if}&\;\;\; X_{m_3}^2 > 0 \nonumber\\
\label{sup1}
\end{eqnarray}
and hence in both cases $\frac{X_{m_3}^2}{v_1 v_2} \leq 0$ is {\sl automatically
satisfied}, so that the ``tree-level logic'' continues to hold at one-loop
in this case. \\

{\bf (1-b) $\tan^2 \beta <1 $}; the proof goes along similar lines as before, 
except that now $a$ and $\tan \beta$ should be of the {\sl same sign}.
In this case, the allowed regions are,
\begin{eqnarray}
1 \geq \tan \beta \geq T_{-} \geq 0 \;\;\; &\mbox{if}&\;\;\; X_{m_3}^2 < 0 \nonumber\\
-1 \leq \tan \beta \leq T_{-}\leq 0 \;\;\; &\mbox{if}& \;\;\;X_{m_3}^2 > 0 \nonumber\\
\label{inf1}
\end{eqnarray}

Nonetheless, there is in this case a difference from the previous one,
namely that the stability condition is not automatic in the region
$-1 \leq \tan \beta \leq 0$ when  $X_{m_3}^2 < 0$ and similarly in the region
$0 \leq  \tan \beta \leq 1$ when  $X_{m_3}^2 > 0$. Both regions are actually 
not excluded by the stationarity condition Eq.(\ref{eqv1v2p}) alone, and one
has to make explicit reference to the stability condition  Eq.(\ref{newcond1}).
Strictly speaking, one should still ask whether the remaining stability
condition Eq.(\ref{eqtbeta}) can guarantee $\frac{X_{m_3}^2}{v_1 v_2} \leq 0$. 
We have actually
exhibited numerical examples where it does not.
Thus in this case, the ``tree-level logic'' {\sl does not go over to the 
one-loop level}.

Now if we compare the two cases of Eqs.(\ref{sup1}) and Eqs.(\ref{inf1}), we 
see that no information is lost if one fixes the sign of $X_{m_3}^2$ from 
the start (by redefining, when needed, the phase of one of the two higgs
fields in Eq.(\ref{vloop}) ). We still insist however, that this choice is 
tricky,
since, in any case, Eq.(\ref{eqtbeta}) is likely to give $\tan \beta$
solutions with opposite signs ( in contrast with the tree-level case
where the two solutions had the same sign, see Eq.(\ref{soltbeta}) ).
It would then be erroneous to choose invariably the positive $\tan \beta$
solutions. Rather, one should keep a handle on the actual sign of 
$X_{m_3}^2$ (corresponding to a given set of values of the MSSM parameters)
and choose the $\tan \beta$ solution which satisfies Eq.(\ref{newcond1}).
Once the right solution is determined, one is at liberty to redefine
its sign and accordingly that of $X_{m_3}^2$. In practice, the wrong choice
would manifest itself through negative $v_1^2, v_2^2$, when
$|\tan \beta| >1$.\\

{\bf Case 2)} $|\tan\beta| \ge \frac{m_t}{m_b}$, ($ \tilde{\alpha} > 0$).
In this case $\tilde{\alpha} ( t^2 - 1) >0$ so that $a(t)$ and $t$ must have
the {\sl same sign}. The allowed regions are then found to be

\begin{eqnarray}
 \frac{m_t}{m_b}\geq \tan \beta \geq T_{+} \;\;\; &\mbox{if}&\;\;\; X_{m_3}^2 < 0 \nonumber\\ 
T_{+} \leq \tan \beta \leq -\frac{m_t}{m_b}   \;\;\; &\mbox{if}&\;\;\; X_{m_3}^2 > 0 \nonumber\\  
\label{infsup2}
\end{eqnarray}

provided of course that 
\begin{equation}
|T_{+}| \leq \frac{m_t}{m_b} 
\end{equation}
Note that, similarly to case {\sl (1-a)}, the stability condition
$\frac{X_{m_3}^2}{v_1 v_2} \leq 0$ is not automatically satisfied through the
sign requirements of $a$ and $t$, and has to be explicitly used
to exclude the regions $\tan \beta \leq  -\frac{m_t}{m_b} $
(resp. $\tan \beta \geq  \frac{m_t}{m_b} $) when $X_{m_3}^2 <0$
(resp. $X_{m_3}^2 >0$). Thus in this case too, the ``tree-level logic''
does not go over to the one-loop level. Finally, the same liberty in the
choice of the sign of $X_{m_3}^2$ noted in case 1) above, holds here
as well.

\newpage
\renewcommand{\theequation}{B.\arabic{equation}}
\renewcommand{\thesection}{B:}
\begin{center}
\section{The method of invariants}
\end{center}
\setcounter{equation}{0}
This method allows to determine simple positivity conditions of the
eigenvalues of the squared mass matrix, {\it without} computing these
eigenvalues. The invariants of the matrix $M$ are defined as the
coefficients of $\lambda$ in the following (finite order) expansion,

\begin{equation}
Det( 1 + \lambda M)= e^{Tr \, ln ( 1 + \lambda M)}\equiv 
1 + \sum_{i=1}^{dim M} Inv_i \lambda^i
\end{equation}

\begin{eqnarray}
Inv_1= tr(M) && \\
Inv_2=\frac{1}{2} (tr(M^2) - tr(M)^2) &&\\
\mbox{etc...} \nonumber &&\\
Inv_{dim M} = Det M && \\
\nonumber
\end{eqnarray}

One can show that a sufficient and necessary condition for all the eigenvalues
to be positive is that all the above invariants be positive. The nicety of
this property is that whatever the size of $M$ is, one can always determine
analytically the positivity conditions of its eigenvalues by computing
straightforwardly powers of traces of powers of M!

\newpage
\begin{center}
{\bf Figure Captions}
\end{center}

{\bf Fig.1:} The typical behaviour of $t_{\pm}$ versus $\zeta\equiv 
\frac{\tilde{\alpha} v_1 v_2}{X_{m_3}^2} $ and in terms of the sign
of $\tilde{\alpha}$. \\

{\bf Fig.2:} An illustration of the typical dependence of the effective 
potential on the gauge-fixing change. Extremal points have the same
value for any $\xi_1, \xi_2, ...$. However, the curve $\xi$ leading
to a gauge-fixing dependent minimum at $M'$ is not allowed.\\

{\bf Fig.3:} The $t_1$ solution of the stationarity equation for $\tan \beta$
as a function of the renormalization scale $\mu_R$. The bounds $\sqrt{t_{\pm}}$
are also shown to illustrate the rejection of $t_1$.\\

{\bf Fig.4:} Same as Fig.3 for the non-rejected solution $t_2$. The tree-level
value $\tan \beta_0$, the relevant bounds $T_{+}$ and $\sqrt{t_{+}}$
and the scales at which the potential is unbounded from below are also
shown.\\

{\bf Fig.5:} Sensitivity of the tree-level and one-loop vev's to the 
renormalization scale. The one-loop vev's have an improved behaviour
in the vicinity of and below the UFB scale.

\newpage
\centerline{\epsfig{file=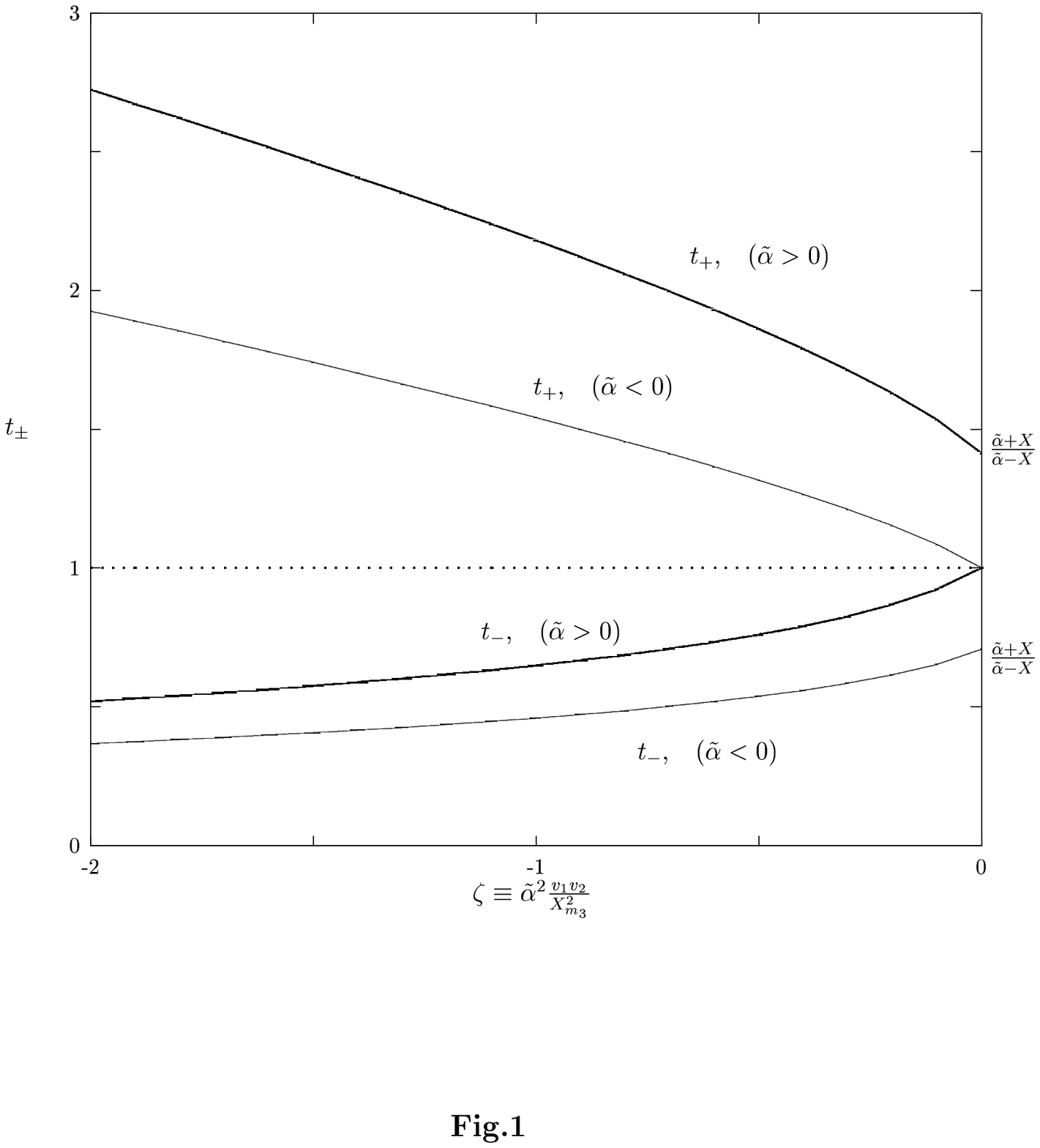, angle=0}} 
\newpage
\centerline{\epsfig{file=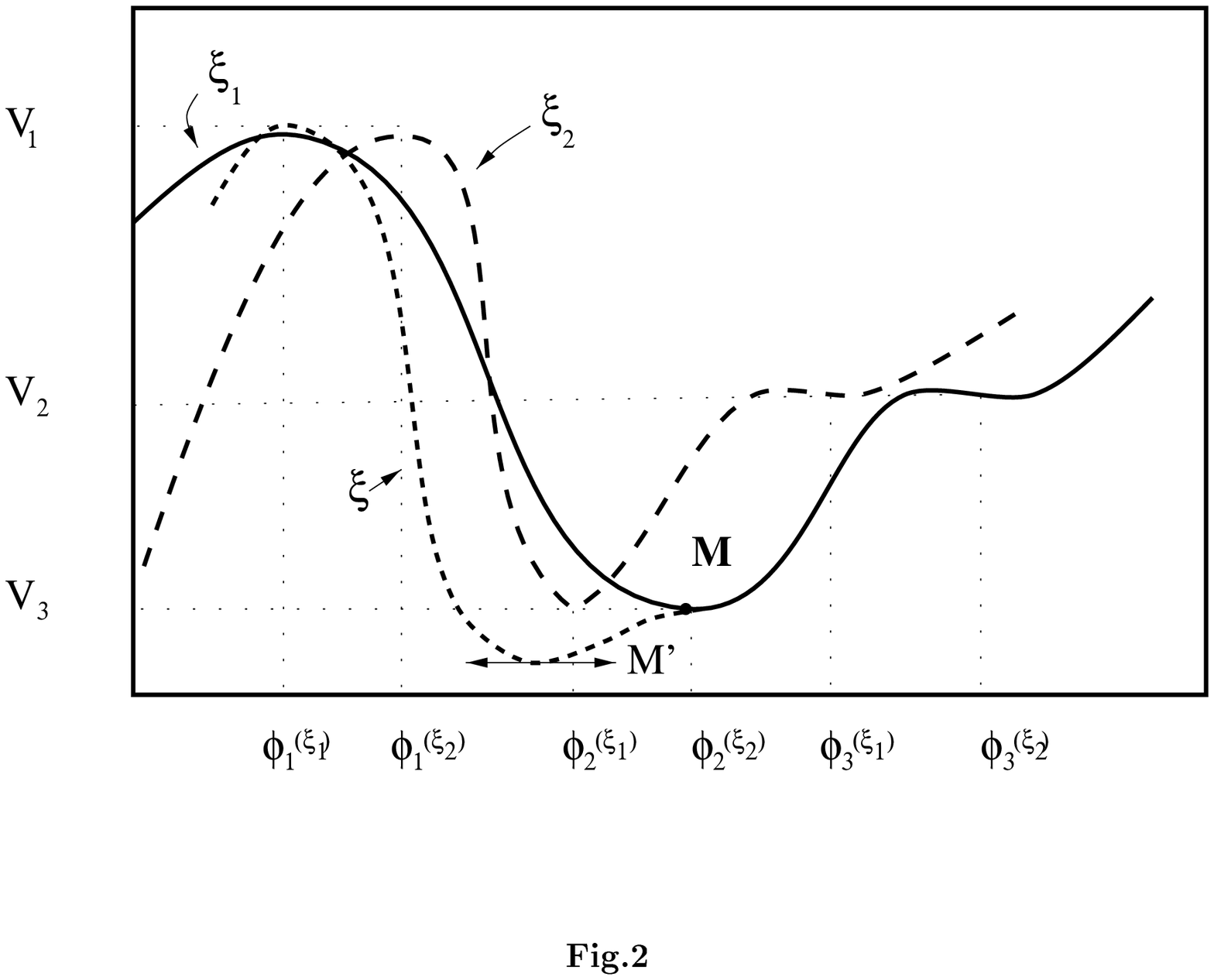, angle=0}} 
\newpage
\centerline{\epsfig{file=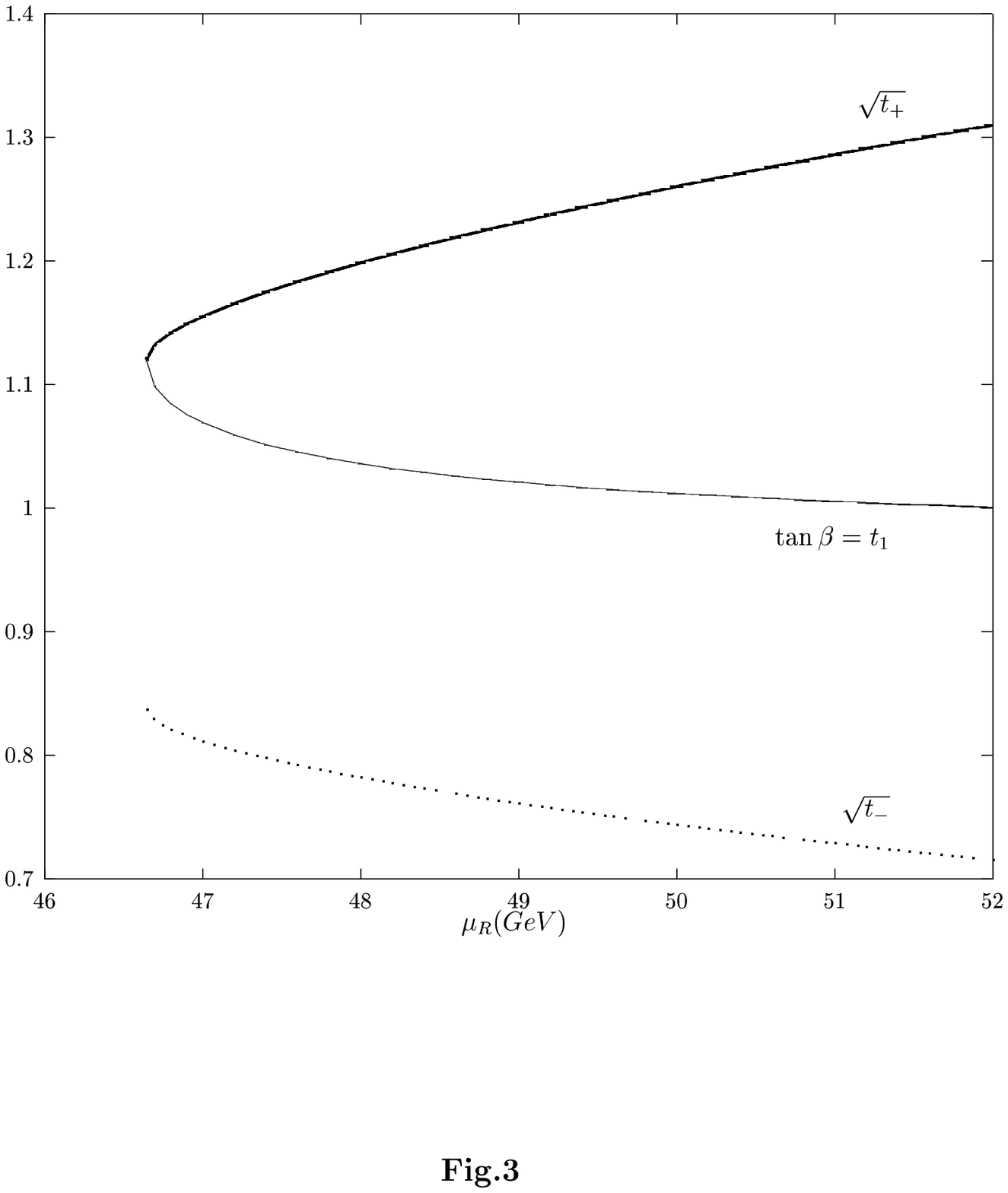, angle=0}} 
\newpage
\centerline{\epsfig{file=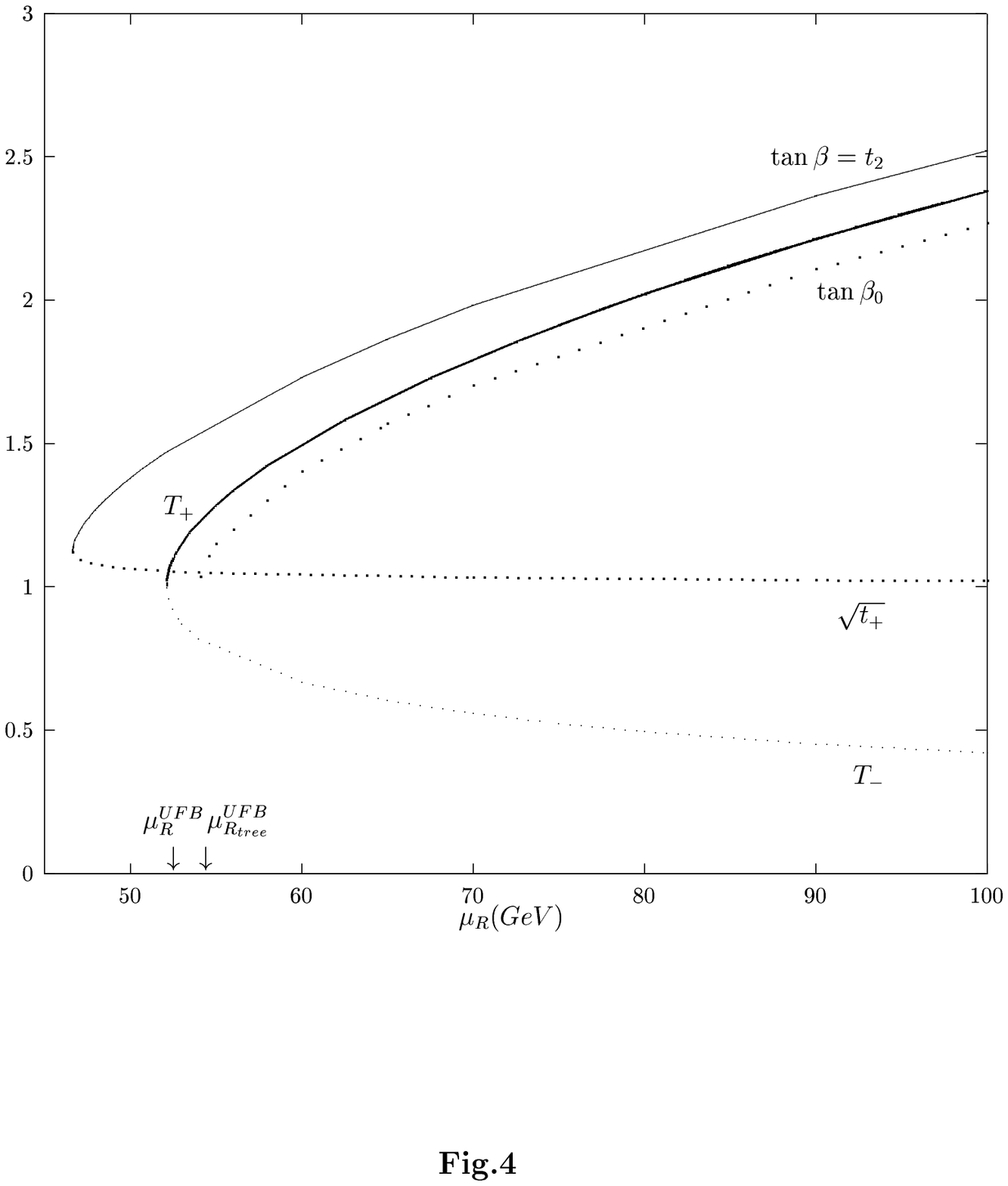, angle=0}} 
\newpage
\centerline{\epsfig{file=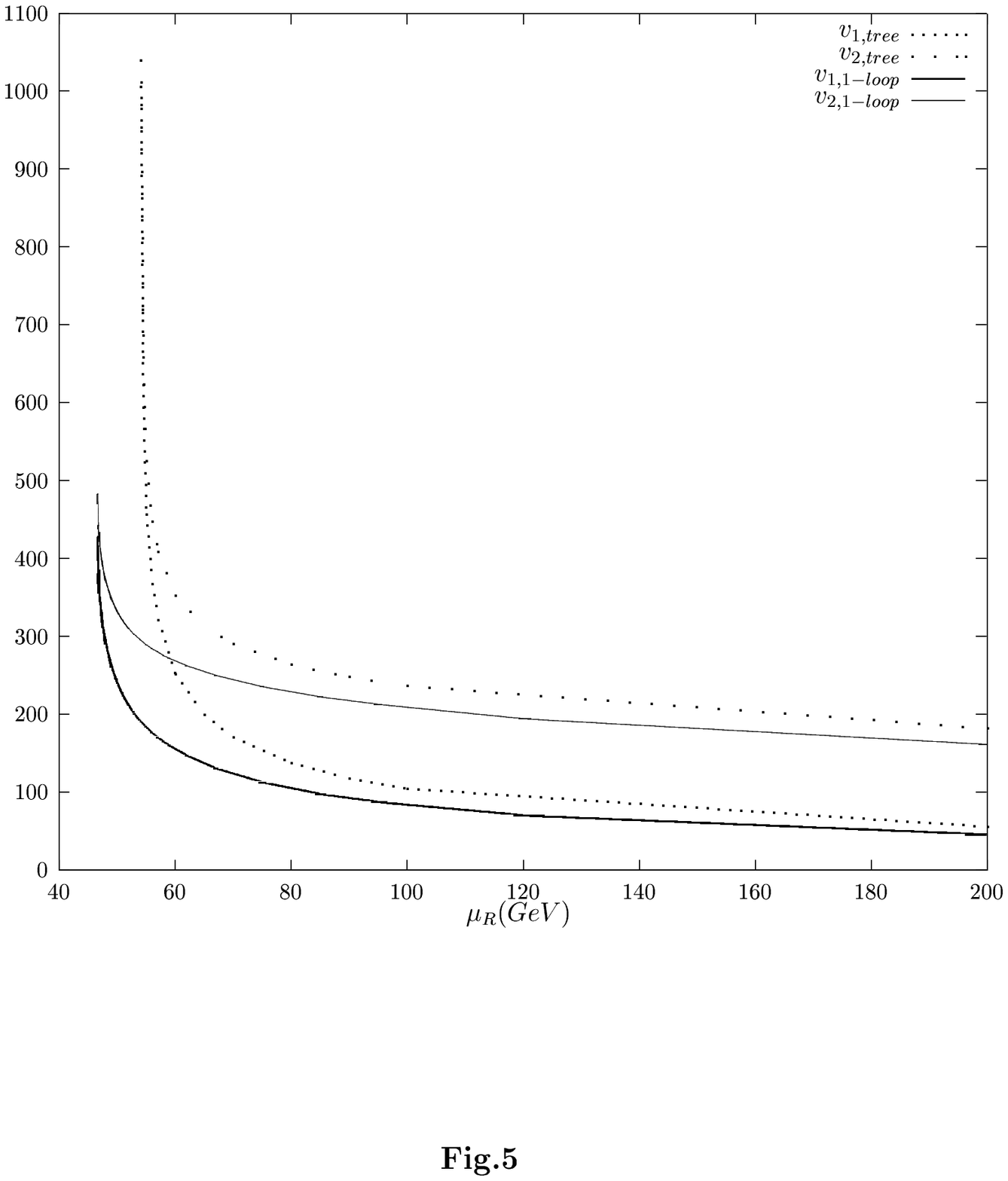, angle=0}}   


\newpage
\begin{thebibliography}{99}
\bibitem{EWB} 
L.E. Iba\~nez and G.G. Ross, Phys. Lett. B110 (1982) 215;
K.Inoue, A. Kakuto, H. Komatsu and S. Takeshita, Prog. Theor. Phys. 68 (1982) 
927; 71 ( 1984) 413;
L. ALvarez-Gaum\'e, M. Claudson and M.B. Wise, Nucl. Phys. B207 (1982) 96;
J. Ellis, D.V. Nanopoulos and K. Tamvakis, Phys. Lett B121 (1983) 123;
L.E. Iba\~nez, Nucl. Phys. B218 (1983) 514; 
L. ALvarez-Gaum\'e, J. Polchinski and M.B. Wise, Nucl. Phys. B221 (1983) 495;
J. Ellis, J.S. Hagelin, D.V. Nanopoulos and K. Tamvakis, Phys. Lett. B125 
(1983) 275; 
L.E. Iba\~nez and C. Lopez, Phys. Lett. B126 (1983) 54; Nucl. Phys. B236 (1984)
438; 
For a pedagogical review, see for instance R. Arnowitt 
and P. Nath, Lecture at Swieca School, Campos do Jordao, Brazil, Jan. 1993, 
and references therein;

\bibitem{hiddensector}
A.H Chamseddine, R.Arnowitt and P.Nath, Phys.Rev.Lett. 49 (1982) 970;
R.Barbieri, S.Ferrara and C.A Savoy, Phys.Lett. B119 (1982) 343;
L.Hall, J.Lykken and S.Weinberg, Phys.Rev. D27 (1983) 2359;

\bibitem{gaugemed}
M.Dine and A.E.Nelson, Phys.Rev.D48 (1993) 1277;

\bibitem{casas} see for instance J.A. Casas, A. Lleyda, C. Mu\~noz,
Nucl. Phys. B471 (1995) 3, and references therein;

\bibitem{spectrum}
D.J. Casta\~no, E.J. Piard and P. Ramond, Phys.Rev.D49 (1994) 4882;
W. de Boer, R. Ehret and D.I. Kazakov, Z. Phys. C67 (1994) 647;
V. Barger, M.S. Berger and P. Ohmann, Phys.Rev.D49 (1994) 4908;

\bibitem{gamb}
G. Gamberini, G. Ridolfi, F. Zwirner, Nucl. Phys. B 331 (1990) 331;


\bibitem{hon}
see \cite{sher} and the first ref. in \cite{spectrum};


\bibitem{moriond}
G. Moultaka, hep-ph/9705330,
(Talk given at the {\it 32nd Rencontres de Moriond: 
Electroweak Interactions and Unified Theories}, Les Arcs, France,
15-22 Mar 1997, to appear in the proceedings.)
and C. Le Mou\"el, G. Moultaka, hep-ph/9708368
(talk given at the EEC Network meeting 
{\it Tests of Electroweak Symmetry Breaking},
Ouranoupolis, Greece, May 27 - June 1, 1997, to appear in the proceedings)

\bibitem{christophe}
C. Le Mou\"el, Ph.D. thesis, in preparation;

\bibitem{decarlos}
B. de Carlos and J.A Casas, Phys. Lett. B309 (1993) 320;
R. Arnowitt and Pran Nath, Phys. Rev. D46, 9 ,(1992) 3981;

\bibitem{multiscale}
C. Ford, D.R.T. Jones, P.W. Stephenson, M.B. Einhorn Nucl. Phys. B 395 (1993) 
17;
M. Bando, T. Kugo, N. Maekawa, H. Nakano, Phys. Lett. B 301 ( 1993) 83 and
Prog.Theor.Phys.90 (1993) 405;

\bibitem{bando}
M. Bando, T. Kugo, N. Maekawa, H. Nakano,
Mod. Phys. Lett. A7 (1992) 3379;

\bibitem{CW} S. Coleman, E. Weinberg, Phys. Rev. D 7 (1973) 1888;

\bibitem{susy} For a review of the MSSM and related topics see
H.~P. Nilles, Phys. Rep. {\bf 110}(1984)1;
H.~E. Haber and G. ~L. Kane, Phys. Rep. {\bf 117}(1985)75;
A. ~B. Lahanas and D. ~V. Nanopoulos, Phys. Rep. {\bf 145}(1987)1;

\bibitem{wessb} Supersymmetry and Supergravity by J. Wess and J. Bagger,
Princeton Series in Physics;

\bibitem{giudice}
G.F. Giudice, G. Ridolfi, Z. Phys. C 41 ( 1988) 447;
see also F. Zwirner in ``Physics and Experiments in Linear Colliders'',
Saariselk\"a, Finland Sept.91, Eds. R. Orava, P. Eerola, M. Nordberg;

\bibitem{ellis1}
J.Ellis, G. Ridolfi, F. Zwirner, Phys. Lett. B257 (1991) 83;
\bibitem{ellis2} 
J.Ellis, G. Ridolfi, F. Zwirner, Phys. Lett. B262 (1991) 477;
\bibitem{comp}
H.E. Haber and R. Hempfling, Phys.Rev.Lett. 66 (1991) 1815;
R. Barbieri, M. Frigeni and M. Caravaglios Phys. Lett. B258 (1991) 167;
Y. Okada, M. Yamaguchi and T. Yanagida, Porg. Theor. Phys. 85 (1991) 1;

\bibitem{habertalk}
see also  H.E. Haber, ``Higgs Boson Masses and Couplings in the
Minimal Supersymmetric Model'' and references therein,
(hep-ph/9707213) to appear in {\it Perspectives on Higgs Physics II}
Gordon L. Kane editor;

\bibitem{tbetaperturb}
V. Barger, J.L. Hewett and R.J.N. Phillips, Phys. Rev. D41 (1990) 3421;
B. Schrempp and M, Wimmer, Prog. Part. Nucl. Phys. 37 (1996) 1;


\bibitem{sher}
M. Sher Phys. Rep. 179, Nos 5\& 6 (1989) 273;
\bibitem{xsindep} R. Fukuda, T. Kugo, Phys. Rev. D 13 (1976) 3469; 
I.J.R. Aitchison, C.M. Fraser, 
Ann. Phys. 156 (1984) 1;

\bibitem{kane}
see for instance, G. L. Kane ``Tests and implications of increasing evidence for
superpartners'', Invited talk at XXXII Rencontres de Moriond, Les Arcs, 
March 1997; hep-ph/9705382;

\end{thebibliography}
\end{document}